\documentclass[prd,floatfix,superscriptaddress,preprintnumbers,nofootinbib,letterpaper]{revtex4}
\usepackage{graphicx}
\usepackage{dcolumn}
\usepackage{bm,latexsym,amsmath,amssymb,amsfonts}
\usepackage[colorlinks]{hyperref}
\usepackage{color}
\usepackage{gensymb}
\usepackage{adjustbox}
\usepackage{scrextend}
\usepackage{enumerate}
\def\apj{Astrophys.~J.}
\def\apjs{Astrophys.~J. Suppl.}
\def\aj{Astronom.~J.}
\def\apjl{Astrophys.~J.~Lett.}
\def\prd{Phys.~Rev.~D}
\def\prl{Phys.~Rev.~Lett.}
\def\plb{Phys.~Lett.~B}
\def\mnras{Mon.~Not.~Roy.~Astr.~Soc.}

\def\AnA{Astron.~Astrophys.}

\def\ARAnA{Ann.~Rev.~Astron.~Astrophys.}
\def\ARNPs{Ann.~Rev.~Nucl.~Part.~Sci.}
\def\cqg{Class.~Quant.~Grav.}

\def\ijmpd{Int. ~J. ~Mod. ~Phys. ~D.}
\def\pr{Phys.~Rep.}

\begin{document}

\title{Measuring the Hubble constant: gravitational wave observations meet galaxy clustering}
\author{Remya Nair}
\email{remya@tap.scphys.kyoto-u.ac.jp}
\affiliation{Department of Physics, Kyoto University, Kyoto 606-8502, Japan}

\author{Sukanta Bose}
\email{sukanta@iucaa.in}
\affiliation{Inter-University Centre for Astronomy and Astrophysics, Post Bag 4, Ganeshkhind, Pune 411 007, India}
\affiliation{Department of Physics \& Astronomy, Washington State University, 1245 Webster, Pullman, WA 99164-2814, U.S.A}

\author{Tarun Deep Saini}
\email{tarun@physics.iisc.ernet.in}
\affiliation{Department of Physics, Indian Institute of Science, Bangalore 560012, India}

\date{\today}

\begin{abstract}
We show how the distances to binary black holes measured in gravitational wave observations
with ground-based interferometers can be used to constrain
the redshift-distance relation and, thereby, measure the Hubble constant ($H_0$).
Gravitational wave observations of stellar-mass binary black holes are not expected to be
accompanied by any electro-magnetic event that may help in accessing their redshifts. We address this
deficiency by using an optical catalog to get the distribution of galaxies in redshift.
Assuming that the clustering of the binaries is correlated with that of the
galaxies, we propose using that correlation to measure $H_0$. We show that employing this method on simulated data obtained for 
second-generation networks comprising at least three detectors,
e.g., advanced LIGO - advanced VIRGO network, one can measure $H_0$ with
an accuracy of $\sim$ 8\% with detection of a reference population of 25
binaries, each with black holes of mass 10$M_\odot$.
As expected, with third-generation detectors like the {\it Einstein telescope}
(ET), which will measure distances much more accurately and to greater
depths, one can obtain better estimates for $H_0$. Specifically, we show that with 25 observations, ET can
constrain $H_0$ to an accuracy of $\sim$7\%. This method can also be used to estimate other cosmological
parameters like the matter density $\Omega_m$ and the dark energy equation of state.
\end{abstract}

\preprint{KUNS-2685, LIGO-P1700098}

\maketitle

\section{Introduction}
\label{sec:intro}
We are living in an era of precision cosmology but a complete understanding of the nature of
dark energy, which is responsible for the observed late-time acceleration of the Universe, still
eludes us \cite{de}. Obtaining accurate and precise distance estimates
to sources at cosmological redshifts is paramount in getting an insight into this mysterious component of the Universe. It is known that
the gravitational wave (GW) signal from inspiraling compact object
binaries can allow for a unique way to measure their luminosity distance with
reasonably good precision, when observed with a sensitive enough GW
detector network~\cite{local2,aghosh,nissanke}. The detection of GW170817 in both GW and electromagnetic (EM) waves was
used to determine the Hubble constant $H_0 = 70^{+12}_{-8}$ km/sec/Mpc \cite{ligo_H0}.

The work we present here
relies on two pillars of observational astronomy: multiple cosmological observations of
the clustering of galaxies and gravitational-wave measurements. We
propose a method to combine them in order to measure cosmological
parameters and to understand the expansion history of the Universe. We
do so after accounting for a selection effect that arises due to the
fact that the sensitivity of a network of GW detectors to inspiraling
binaries varies with their sky-position, orbital inclination, and distance.

Type Ia supernovae (SNeIa) are termed as `standard candles'  in cosmology; distance
estimates obtained from their observations helped to map the cosmic expansion history of the
Universe. Since different models of the Universe may predict somewhat different evolution, these
distance estimates can even be used to test the validity of these models. For robust tests of these 
models it is desirable to have a variety of observations that probe the observable Universe
in different ways. This helps in identifying and mitigating systematic
errors. 

In the past, measurements
from a number of observations have been employed, either independently or in combination, to get
estimates of cosmological parameters. These observations include (but are not limited to):  SNeIa,
baryon acoustic oscillations (BAO), galaxy ages, cosmic microwave background (CMB), weak
lensing, etc. \cite{rev_wein,de_obs}. SNeIa measurements gave the first persuasive evidence for
the existence of the recent accelerated phase of the expansion of the Universe \cite{sneia}. Future
surveys, like the Dark Energy Survey (DES), \cite{des}, the Panoramic Survey Telescope and Rapid
Response System (Pan-STARRS) \cite{pans}, and the Large Synoptic Survey Telescope (LSST)
\cite{lsst} will substantially increase the number of SNeIa candidates, but the estimates on
cosmological parameters obtained from these may be limited by the
systematic errors. 

With so many new SNeIa discoveries, it is expected that soon the systematic uncertainties will become
comparable to statistical uncertainties owing to the diminishing value of the latter with every new observation 
\cite{sneia_syst1}. Further, it is now also understood that
systematic uncertainties in SNeIa observations are correlated. Potential sources of systematics
include variations of SNeIa magnitudes that correlate with the properties of their host galaxies \cite{sneia_syst2},
and model assumptions in the light-curve fitting methods used to standardize the SNeIa candidates \cite{sneia_syst3}. 
An additional problem is that distance estimates obtained from SNeIa observations are
indirect and depend on a distance ladder, where measurements of nearby stars are used to
calibrate distances to far-off objects in a series of steps (see \cite{hst} and references therein). Any uncertainties in such a calibration can
add significant errors to the distance estimate of sources at large (cosmological) distances.

Another cause of concern in finding robust constraints on cosmological parameters is the
inconsistency between the parameter estimates from different cosmological probes. For example the
constraints on the Hubble constant from non-local experiments, like the CMB
measurements of the Planck satellite \cite{planck14}, have been in significant tension with the
results of the Hubble space telescope (HST) \cite{riess11}. The latest Planck High Frequency
Instrument (HFI) data \cite{planck16}, confirmed and further increased this tension. Its latest
estimate of $H_0$ is 66.93 $\pm$ 0.62 km/sec/Mpc (68\% confidence
level) against the improved HST estimate of 73.00 $\pm$ 1.75 km/sec/Mpc (68\% confidence
level) \cite{riess16}. Hence the inconsistency in the estimate of the Hubble constant now stands
at a staggering $>$ 3$\sigma$ confidence level (assuming standard $\Lambda$ cold dark matter (LCDM) cosmology).
The tension can be somewhat abated by considering an extended LCDM scenario and
allowing for phantom equation of state for dark energy, but the tension resurfaces upon the addition
of BAO or SNeIa data \cite{valentino}. Moreover, disagreement between distance estimates from
BAO and SNeIa measurements have been reported in the past \cite{sneia_bao}. In such a scenario,
having a new observational window to view the Universe is very appealing.

The detection of gravitational waves by the LIGO-VIRGO detector network \cite{gw150914,gw150914properties,gw150914calibration,gw_det} have ushered in the era of GW astronomy. Analogous to standard-candle
SNeIa, GW measurements of inspiraling neutron star (NS) or black
hole (BH) binaries can be used as `standard sirens' in cosmology \cite{schutz}. GW measurements
will provide independent distance estimates that will complement other probes of
precision cosmology mentioned earlier. Since the underlying assumptions, the
observational techniques, the biases and the systematic errors of all these
probes are decidedly different, the hope is that requiring consistency among
them and combining them for parameter estimation will help in identifying
systematic errors and model dependencies. The idea of using GW sources as
standard sirens in cosmology, was initially put forward by Schutz in 1986, where
it was shown that kilometer-sized GW interferometers can be used to constrain
cosmological parameters, like the Hubble constant, to an accuracy of 3\% using
observations of coalescing NS binaries \cite{schutz}. GWs from
compact binary mergers can give physically calibrated absolute distances to
sources at large redshifts, i.e., unlike SNeIa measurements these do not depend
on a distance ladder. The calibration here lies in the assumption that general
relativity accurately describes the gravitational waveform. The measurement of
the wave amplitude, the frequency, the chirp rate (rate of change of frequency)
and the orbital inclination angle of the binary system, from a network of GW
detectors, contain information about the luminosity and brightness of the GW
source. The chirp rate is a measure of the luminosity of the compact binary system
since the change in frequency is caused
by the energy loss through emission of GWs, and the observed amplitude is a
measure of the brightness. Therefore the luminosity distance to the
faraway source can be inferred from these observations. 

Although the GW signals
provide a direct measurement of distance, they do not provide a redshift estimate
of the source, which is essential if one wants to constrain the distance-redshift
relationship in cosmology. The scale-invariance of the binary black hole (BBH) 
waveforms with redshifted mass implies that GW
signals from a local compact binary with component masses ($m_1,m_2$) would be
indistinguishable from the GW signal from a compact binary at a redshift $z$ with
component masses ($m_1/(1+z),m_2/(1+z)$). Hence, to use these distances in
cosmology one requires an independent measure of 
redshift. Identifying the host galaxy is one way this 
information can be obtained. That would, however, require
good localization of the GW source in the sky.

The projected sky localisation accuracy of a three-detector network
comprising ground-based detectors of Advanced LIGO (aLIGO) and
Advanced VIRGO (AdV)~\cite{ligo,virgo}, operating at their respective
design sensitivities, can range from about a tenth of a sq. deg. to a
few tens of sq. deg. with median value of a few sq. deg. (at 68\%
confidence) for a BBH with total mass of $20M_{\odot}$ and at a luminosity
distance of 1~Gpc~\cite{gw150914,gw_det,local1,local2,aghosh}.
Thus, a few square-degrees is a reasonable estimate for localisation
accuracy of a second-generation three-detector network for a range of source masses
similar to those of the BBHs already observed in GWs. Since such a sky patch can contain thousands of galaxies, 
the chances of identifying the host galaxy of a GW event from a galaxy
catalog can be dismal. 

Note that the latest NS-NS discovery by the LIGO-VIRGO network was accompanied by an extensive
follow-up effort by around 70 ground- and space-based observatories which observed the sky localisation
patch given by the GW signal across the electromagnetic spectrum. This
led to the discovery of the first EM counterpart to a GW event \cite{gw_em}. It is difficult to say at this time whether there would be many
such NS-NS events in the future or if we just got lucky this time (given that the event was very near at $\sim41$ Mpc).
For the foreseeable future these kind of EM-counterpart detections will be limited to low redshifts. 
Nevertheless, the angular resolution is expected to get better in the future with the addition of KAGRA 
\cite{kagra} and LIGO-India \cite{ligo-I} in the network, but even then the probability 
of identifying the host galaxy for far redshift sources unambiguously,
may remain small. 

The main aim of this work is to show how the distances
measured from GW observations
of binary black holes with ground-based interferometers can be used to constrain
the redshift-distance relation and hence estimate the Hubble constant.
After showing how this idea can be implemented in a realistic scenario
with the aLIGO-AdV three-detector
network, we also demonstrate the improvement in accuracy that can be
achieved  with observations of the same systems in the third-generation
observatory in the form of the Einstein Telescope (ET)~\cite{et}.
We begin in the next section with a synopsis of a variety of
approaches that have been applied in the past to determine the
redshift of the GW standard candles.

\section{Proposals for obtaining cosmological parameters from GW sources} 

Various methods have been proposed to constrain cosmological models using GW signals from coalescing binaries. 
We recapitulate the ones most relevant to our method here.

\subsection {Electro-magnetic counterpart to the GW event}
The most straightforward way to measure the redshift associated with a GW
event is to identify its EM counterpart.
Gamma ray bursts (GRBs) are powerful beams of radiation lasting a few seconds,
and are mainly classified into two categories: long GRB that lasts for more than
two seconds and short GRB that typically lasts for less than two seconds. The
nature of the short GRB has been disputed for very long \cite{sgrb_NS} but the observation
of GRB 170817A  has confirmed binary neutron star mergers as a progenitor of short GRBs~\cite{EM_gw}.
The assumptions of evolutionary processes on the formation of NS-NS/NS-BH compact
binaries, the metallicity models, the star formation rates, etc., all have notable
effects on the estimates of the merger rates of these compact binaries. Hence the
merger rates have fairly uncertain theoretical estimates and they are poorly
constrained from observations. For example the detection rates for
binary NS mergers were projected to range from 0.4 to 400 events per
year with advanced LIGO (aLIGO) design sensitivity in
Ref.~\cite{rates_abadie}, but might reduce following the observation
of GW170817. For BBH sources the detection rates in the same reference
were projected to be in the range from 0.4 to 1000 events per year,
but have been revised to be in the (narrower) range from 2 to 600 Gpc$^3$yr$^{-1}$~\cite{Abbott:2016nhf}. 
Although EM counterparts are not expected from BBH binary
mergers, NASA's Fermi telescope detected a GRB 0.4 seconds after 
GW150914. The GRB lasted for 1 second \cite{fermi_grb} and is possibly 
not connected  with the GW source \cite{loeb,integral}. 

The EM follow up of GW inspiral events is a challenging
task~\cite{em_follow}. What adds to the demanding exercise of
detecting a possibly highly beamed and short-lived
signal, is the contrast between the sky localization accuracy of
current GW networks and the fields of view of optical telescopes. The
localisation provided for the GW events observed so far is poor,
{\it viz.} $\sim$100s of sq.~deg. (the sky localisation for the first GW detection had an area of ~600~sq.~deg.).
Comparing this with the fields of view of optical telescopes like the Zwicky
Transient Facility (47 sq.~deg.), the Dark Energy Camera (3 sq.~deg.) and the LSST
(9.6 sq.~deg.) gives an idea about the formidable challenge faced by astronomers in
following up these events. Since the EM signals could be short lived and may peak
within hours (or faster), successful EM follow up would require accurate sky
localizations within a time scale of minutes to hours. To this end, many algorithms
have been developed or are in development to account for telescope pointing limitations,
finite observation time, the rising or setting of the target at the observatory location,
etc.~\cite{em_follow}. As mentioned in the introduction, the
LIGO-VIRGO network has already detected an event, GW170817, with 
confirmed EM counterparts~\cite{ligo_H0,gw_em}. Moreover, future GW
networks will have narrower sky localization regions, as mentioned
above. However, it is still premature to say at this time whether there would be many
such NS-NS events in the future.
Nevertheless multiple studies have been performed assuming
simultaneous observations of the GWs and EM signatures to constrain cosmological
parameters like the Hubble constant \cite{gw_em_cosmo}.

\subsection{Neutron star mass distribution}
The knowledge about the intrinsic mass distribution of the NS population can also be
used to estimate source redshifts \cite{markovic}. The GW signals give an estimate of
the redshifted mass of the binary $m_z=m(1+z)$ and if the distribution of NS masses
is known, one can obtain a distribution of the source redshift. The number of detected
pulsar binaries have steadily increased in recent years and current observations
estimate that the mass distribution of NS in binary NS systems could be multi-modal where
the two modes in the distribution are expected to be associated with different NS
formation channels. 
The idea has been explored in multiple publications \cite{gw_ns} but depends
on the knowledge of NS mass distribution and may be prone to systematics from selection
biases. To complicate things further, the mass distribution in double NS systems may be
different from that in other systems with a single neutron star \cite{ns-mass}.

\subsection{Tidal deformation of the neutron star}
The correction to the GW waveform due to the finite size of the compact object in a
binary system depends on the equation of state as well as the rest masses. If these
corrections can be measured from GW signal they will provide information not just about
the redshifted mass but also the rest mass of the system, hence providing an estimate
of the redshift. The idea has been explored in the literature and it was shown that
as small as 10\% error on redshift estimate can be expected \cite{gw_tidal}. Here too
the analysis depends on the knowledge of equation of state of the NS, which is highly
model dependent and hence prone to systematic errors.

\subsection{ Statistical techniques with galaxy catalogs}
The technique that is closest to our approach in this work is the use of existing
galaxy catalogs. Schutz \cite{schutz} proposed the use of sky position-luminosity distance
confidence regions informed from GW measurements and statistically ruling out galaxies 
that did not host the event. The method has been modified and developed further in a
number approaches, e.g., a Bayesian framework that incorporates assumptions and prior
information about a GW source within a single data analysis framework \cite{walter},
using clustering of galaxies to statistically extract the redshift information from
a GW sample without identification of host galaxies for individual events \cite{hogan},
and using sources with known redshifts to iteratively solve for the redshift of
unresolved sources \cite{tsv}. The method we present here can be considered as a natural extension of 
\cite{walter,hogan}

\section{Methodology}\label{sec:method}

Stellar-mass BBHs are expected to reside in galaxies or their
neighborhood. Therefore, we assume that their spatial distribution
follows that of galaxies. If the galaxies were uniformly randomly
distributed in volume, then these sources of GWs~\footnote{In this 
	work, by GW sources we always mean BBH, and by BBH we always imply a binary 
	of stellar-mass black holes that is a source of GWs for ground-based detectors. For obtaining
	the results we set the mass of each black hole in every binary to
	10~$M_\odot$. We leave it for future
	work to study how different black hole mass distributions affect these results.}
would also have been distributed in the same manner. In that case, due to the
absence of spatial features in the distribution of GW sources and
galaxies, the two point-sets would be uncorrelated. In reality, due to
gravitational instability, galaxies show strong clustering on
length-scales below about 100 Mpc. The spatial clustering of galaxies
in the ($z, \theta, \phi$) space would  then correspond to a
distribution of GW sources in the ($D_L, \theta, \phi$) space.
Therefore, while matching the patterns in the two distributions --- assuming they are
coincident, since BBH sources are located in and around galaxies --- a 
method to associate $z$ with $D_L$ can be obtained. This would lead to the distance-redshift relation for such objects. However, it is clear that if the angular location of
BBHs as  GW sources is not narrow enough, then one is compelled to sample the angle-averaged galaxy clustering. Owing to the angle averaging the distribution now depends only on the redshift of the galaxies (or equivalently on $D_L$ for GW sources). This blunts the impact of the galaxy clustering owing to the overlapping in the redshift space of clustering in different directions. The important thing to note is that the near-future GW experiments have an angular resolution, which although not enough to single-out an individual galaxy as the source of a GW event, is nonetheless sharp enough to probe the clustering length of about 100 Mpc in the nearby universe~\cite{Chen:2016tys}.

\begin{figure*}[!htbp]
	{\centering
		
		\includegraphics[height=0.33\textwidth]{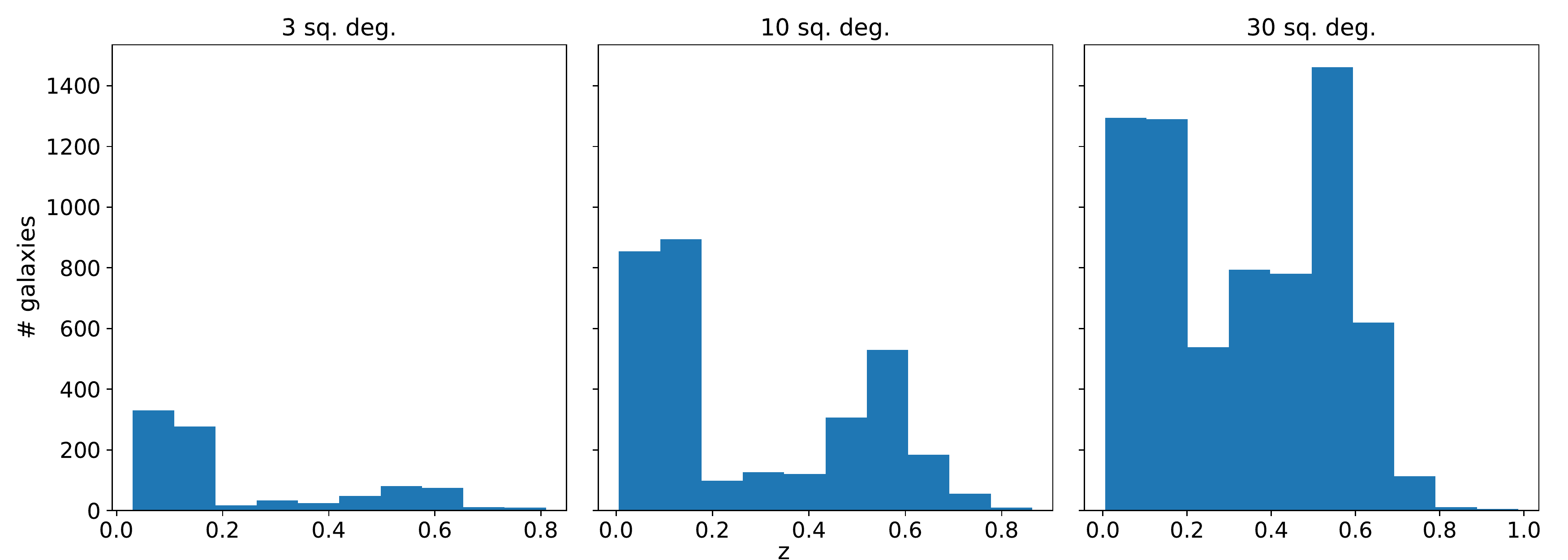} \label{zDista}	
		\caption{The redshift distribution of galaxies in three random sky patches from the SDSS catalog.}
		\label{zDist}
	}
\end{figure*}

To explore the efficacy of this idea, we need to simulate GW sources taking into account their spatial clustering. Since we are assuming that the sources reside in and around galaxies, this can be done most effectively by using galaxy redshift catalogs. For our purpose we make use of the existing galaxy catalog: the Sloan Digital Sky Survey (SDSS)
\cite{sdss}. In Fig. \ref{zDist} we show the redshift distribution of
three random SDSS patches of 3, 10 and 30 sq. deg.. We chose these regions randomly from the sky area covered by SDSS and we have not employed any mass (luminosity) cuts on the galaxies in this work. 
The galaxy distribution will change if the patches are narrowed 
to a tenth of a square degrees or broadened to a few hundreds of square degrees. 
Very narrow and very broad patches will reduce the power of the
method described here since both types will limit the optimal 
use of galaxy clustering information (more about this below).
We expect our qualitative results to hold as long as the sky patches are no larger than 30 sq. deg.. We further emphasize that these numbers do not represent the `optimal' sky patch area. A much detailed analysis can be carried out to study the effect of patch size on the estimates. We leave this exercise for future. 
Although the SDSS is not a complete all-sky catalog, it
is wide enough to sample several sky patches of the size that typical
angular resolution for a GW source necessitates.  We are also constrained by the sensitivity of the near-future GW experiments, and the limiting redshift probed by the SDSS falls neatly in that range. The formalism described here would require a more complete all-sky galaxy redshift catalog for use in the near-future experiments than what exists today.    

\subsection{Bayesian formulation}

The goal of this section is to formulate a method for extracting cosmological information from GW sources when their redshifts are not known. We do this by first constructing the Probability Distribution Function (PDF) for cosmological parameters in the most general case. We subsequently use reasonable simplifications to produce an approximate PDF that is used to test the efficacy of our method.

The spatial location of a GW source, given by its angular position and
luminosity distance, can be determined only up to a limited precision
that depends on its redshift, GW luminosity, orientation, sky
position, and the detector network~\cite{local2}. The measured angular
position of the source and its luminosity distance would in general be
given by a likelihood function 
that is obtained from an analysis of GW data by
marginalizing all parameters other than those that determine the
location of the source in three-dimensional space. 
Broadly speaking, for loud enough BBH events as GW sources this distribution can be expected to peak in a certain direction $\Omega_0 \equiv
(\theta_0, \phi_0)$ and at a certain luminosity distance $D_L = D_0$,
and then fall-off away from the peak roughly at a rate that depends on
the sensitivity of the  detector configuration and the nature of the
source and its physical parameters. 

To infer cosmological parameters from GW observations one needs to
relate them to such a likelihood function. Given the choice of cosmology, the source redshift $z_s$, and source angular coordinates $\Omega_s$, the likelihood function can be expressed as 
\begin{equation}
P({\cal S}\,|\, z_s,\Omega_s, \boldsymbol \xi) = {\cal L} \left(\Omega_s, D_L(z_s, \boldsymbol \xi)\right)\,,
\label{eq:likelihood}
\end{equation}
where $\boldsymbol \xi$ represents the parameters of the cosmological model, and ${\cal S}$, as a shorthand for (standard) siren, represents the GW data. 

If the location of the source $(z_s, \Omega_s)$ is known, the likelihood function is sufficient to constrain the cosmology, although we would need a large number of GW sources to better constrain cosmological parameters $\boldsymbol \xi$. Assuming that the GW signal is not accompanied by an electromagnetic counterpart, this information ($z_s$) would in general not be available to us. However, if we assume that the GW source originates inside a galaxy (hitherto unknown to us), then the least we know is that the source position would coincide with the location of that galaxy. In fact, without any more information at hand, we know \emph{only} that the source resides in \emph{one} of the galaxies in the universe. 

We proceed by assuming that all the galaxies in the universe are equally likely to host the source of our GW signal. If we know the redshift and angular position of \emph{all} the galaxies in the sky, the Bayesian prior PDF for the source parameters $(z_s, \Omega_s)$ can be written as   
\begin{equation}
P_s(z_s,\Omega_s) \propto \sum_i \delta(z_s-z_i) \delta^2 (\Omega_s - \Omega_i)\,,
\label{eq:delta} 
\end{equation}
where the sum is over \emph{all} the galaxies in the universe, and the omitted normalization constant is $1/N_g$, where $N_g$ is the total number of galaxies in the Universe. Note that although this is formally correct, in practice we need only those galaxies that are roughly in the direction of the source, as we argue in a following section (\S~\ref{sec:approx_likelihood}). The two-dimensional delta function can be written explicitly as   
\begin{equation}
\delta^2 (\Omega_s - \Omega_i) = \frac{\delta(\theta_s - \theta_i)\delta(\phi_s - \phi_i)}{\sin(\theta_s)}\, ,
\end{equation}
where $\theta$ and $\phi$ are the usual spherical polar coordinates to locate galaxies on the celestial sphere. The denominator in this expression comes about since this is probability density per unit solid angle. If the galaxy redshifts are not known precisely but contain Gaussian noise, and the angular positions of galaxies are known precisely, then we can write $P_s$ as
\begin{equation}
P_s(z_s,\Omega_s) \propto \sum_i \exp\left [-\frac{(z_s-z_i)^2}{2\sigma_i^2} \right] \delta^2(\Omega_s - \Omega_i)\,,  
\label{eq:gaussian}
\end{equation}
where $\sigma_i$ is the error in the redshift of the $i^{\rm th}$
galaxy. Even when this error is small, it is useful to choose the 
Gaussian spread to be somewhat large since it helps to make the
discrete galaxy distribution in to a more continuous one and, thereby,
help in a more meaningful correlation with a GW source distribution. We discuss this point more in \S \ref{sec:num_sim}.
We have omitted the normalization constant
above, and shall continue to do so below. 

Without any other available information we can assign this as the
prior PDF for $(z_s, \Omega_s)$. However, not all galaxies are equally
likely to host GW events; in fact, we expect the probability that a
certain galaxy is the source of our GW signal to be at least
proportional to its mass, or its type (spiral or elliptical). Also, if
the detector configuration is insensitive to GW sources beyond a
certain distance, we can use this information to further reduce the
number of galaxies required for the construction of our prior PDF
$P_s(z_s,\Omega_s)$. Therefore, in general, we modulate the prior
distribution $P_s$ with a weight function $W_i$ affixed to the $i$th
galaxy inside the summation sign to take into account additional
astrophysical/detector information. This weight function determines
the likelihood of the $i^{th}$ galaxy to be the host of the GW source. We
show this explicitly in the next section. 

If the prior PDF for the cosmological parameters is $P_c(\boldsymbol \xi)$, then the complete prior joint PDF is given by
\begin{equation}
P(z_s,\Omega_s,  \boldsymbol \xi) = P_s(z_s,\Omega_s) P_c(\boldsymbol \xi)\,.
\end{equation}
Using the Bayes theorem we can now write the posterior PDF for
$(z_s,\Omega_s, \boldsymbol \xi)$ in terms of the likelihood function and the prior PDF as
\begin{equation}
P(z_s,\Omega_s, \boldsymbol \xi\,|\,{\cal S}) \propto P({\cal S}\,|\, z_s,\Omega_s, \boldsymbol \xi) P(z_s,\Omega_s, \boldsymbol{\xi})\,.
\end{equation}
After marginalizing this over the source parameters $(z_s,\Omega_s)$ we obtain
\begin{equation}
P(\boldsymbol \xi\,|\,{\cal S}) \propto \int dz_sd\Omega_s P({\cal S}\,|\, z_s,\Omega_s, \boldsymbol \xi) P(z_s,\Omega_s, \boldsymbol{\xi})\,.
\end{equation}
For the prior source distribution given by Eq.~\eqref{eq:delta}, the integrals can  be done analytically to give
\begin{equation}
P(\boldsymbol \xi\,|\,{\cal S}) \propto \sum_i {\cal L} \left(\Omega_i, D_L(z_i, \boldsymbol \xi)\right) P_c(\boldsymbol{\xi})\,,
\end{equation}
and if the prior source distribution is given by Eq.~\eqref{eq:gaussian} then we obtain
\begin{eqnarray}
\nonumber
P(\boldsymbol \xi\,|\,{\cal S}) \propto \int &dz_s& \sum_i \Bigg(\exp\left [-\frac{(z_s-z_i)^2}{2\sigma_i^2} \right]
 {\cal L} \left(\Omega_i, D_L(z_s, \boldsymbol \xi)\right) P_c(\boldsymbol{\xi}) \Bigg),
\end{eqnarray}
where the integral over redshift can be converted to an integral over
distance. Henceforth, we work with this posterior.

Till now in our formulation we have considered only a single GW source. Noting that the posterior PDF in the
last equation $P(\boldsymbol \xi\,|\,{\cal S})$ can be used as a prior PDF for another source, we can easily combine data from different GW events. Formally, if $\{{\cal S}_{\rm new}\} = \{{\cal S}, {\cal S}_{\rm old}\}$ then
\begin{eqnarray}
\nonumber
P(\boldsymbol \xi\,|\,{\cal S}_{\rm new}) & \propto & \int dz_s \sum_i \Bigg(\exp\left [-\frac{(z_s-z_i)^2}{2\sigma_i^2} \right]
 {\cal L} \left(\Omega_i, D_L(z_s, \boldsymbol \xi)\right) P(\boldsymbol \xi\,|\,{\cal S}_{\rm old}) \Bigg)\,.
\end{eqnarray}
Note that ${\cal S}_{\rm old}$ contains all the GW sources analyzed till the point the new source $\cal S$ is added to create the updated data set ${\cal S}_{\rm new}$. This recursion can be easily used to obtain a single expression for the combined data, but is notationally somewhat cumbersome. To investigate the efficacy of this method using  simulated data, however, we make simplifications by approximating the likelihood function as an error-box in the following section.

\subsection{Approximate likelihood function}
\label{sec:approx_likelihood}
Note that although in our formulation, thus far, the sum is over \emph{all} the galaxies in the sky, in practice the dominant contribution comes only from the galaxies that are in directions where the likelihood function is significant. Moreover, if prior information is also assumed for the cosmological parameters, a rough measure of the redshift of the source is then known, and only galaxies with redshift close to that value would contribute to this sum. To see this more clearly, let us assume that our likelihood function is approximately given by
\begin{eqnarray}
\nonumber  
{\cal L} (\Omega, D_L) \propto &\exp \left [ -\frac{\displaystyle(D_L-D_0)^2}{\displaystyle 2\sigma_D^2} \right ]&\,, ~{\rm for}~\Omega \in \Delta \Omega\,,\\
= &0&\,, ~{\rm for}~\Omega \notin \Delta \Omega\,.
\label{eq:likelihoodexample}
\end{eqnarray}
Here we have assumed that the measured luminosity distance $D_L = D_0$ with standard deviation  $\sigma_D$, and the angular location of the source is somewhere inside the solid angle $\Delta \Omega$, with all directions inside this solid angle being equally likely. Since the angular position  $\Omega$ is not explicitly present in the likelihood function, the posterior distribution  for $\boldsymbol \xi$ takes the form
\begin{eqnarray}
\nonumber
P(\boldsymbol \xi\,|\,{\cal S}) \propto \int dz_s && \Bigg( \exp \left [ -\frac{\displaystyle(D_L(z_s,\mathcal{\boldsymbol \xi})-D_0)^2}{\displaystyle 2\sigma_D^2} \right ] 
 \times \; n_s(z_s) P_c(\boldsymbol \xi) \Bigg )  
\label{eq:approx_post}
\end{eqnarray} 
with 
\begin{equation}
\label{weight_prob}
n_s(z_s) \propto \sum_{\Omega_i \in \Delta \Omega} W_i \exp\left [-\frac{(z_s-z_i)^2}{2\sigma_i^2} \right]\,, 
\end{equation}
where $W_i$ is a weight function that may be chosen to be the
likelihood of the $i^{th}$ galaxy to be the host of the GW source. For example, this may
depend on galaxy type, galaxy mass or other astrophysical parameters
that determine the rate of GW events in these galaxies. One also needs to
account for detector characteristics: some events will be easier to
observe than others because, e.g., they are located at a nearby
redshift, have an optimal sky position, or have their orbital plane
favorably oriented. On the other hand, this function, or even the full
prior, may be chosen to be uninformative. For the illustration of
our method below, we opt to use an informed prior.


\subsubsection{On the completeness of galaxy catalogs}

It is worth mentioning that although till this point our formulation seemingly requires a complete catalog of galaxies in the universe, it is in fact not necessary. This can be argued in general, but it is easier to argue from the point of view of this approximate formulation as follows. The redshift information required to determine cosmology from luminosity  distance is encoded in the source function 
$n_s(z_s)$. If this function is featureless, then essentially we gain no useful knowledge about the redshift of the source by knowing this function. However, since galaxies are strongly clustered on various length-scales, the distribution of galaxy in space is full of features, such as peaks and troughs in redshift arising due to clusters and voids.  If the sample of galaxies is not complete, i.e., it does not contain \emph{all} the existing galaxies in a given direction, but still captures the dense regions in sufficient detail, then the informative content of galaxy clustering in $n_s(z_s)$ can suffice to constrain cosmology, and in fact can be looked at as pattern matching between galaxy clusters and GW source clusters. However, for the truest match one should use the most complete galaxy catalog available.

\section{Numerical simulations}
\label{sec:num_sim}

To test the efficacy of this method we simulate GW data by using the projected GW source configurations for near-future experiments using the distribution of galaxy redshifts from SDSS. For simplicity, we use the  approximate formalism of \S~\ref{sec:approx_likelihood}. Since the redshift depth of the current experiments is likely to remain shallow, at the most such observations will be able to constrain the Hubble constant. The luminosity distance for a flat LCDM model is given by
\begin{equation}
D_L(z) =  \frac{c(1+z)}{H_0} \int_0^z \frac{dz'}{\sqrt{\Omega_m(1+z')^3+(1-\Omega_m)}}\, ,
\label{dl}
\end{equation}
where $\Omega_m$ is the matter density, $c$ is the speed of light,  and $H_0$ is the Hubble constant. For our analysis we fix the matter density to the input value $\Omega_m=0.3$ and do parameter estimation only for $H_0$. 

The analysis in the previous section was carried out in the redshift space. We find it convenient to do our calculations in the distance space.
For this purpose, note that in Eq.~\eqref{eq:approx_post}, the combination $n_s(z_s) dz_s = dN$ is proportional to the number of galaxies in the redshift interval $dz_s$, which is a pure number. This combination can be expressed in the distance space through $n_s(z_s) dz_s = n_s(z_s(D))dD$, where $D = D_L(z_s, \boldsymbol \xi)$. Therefore, for a given cosmology, we translate the galaxy redshifts to luminosity distance using Eq.~\eqref{dl} and construct the number density function in the distance space. Therefore, Eq.~\eqref{eq:approx_post} is modified to
\begin{eqnarray}
\nonumber
P(H_0\,|\,{\cal S}) \propto \int dD && \Bigg( \exp \left [ -\frac{\displaystyle(D-D_0)^2}{\displaystyle 2\sigma_D^2} \right ]  \times \; n_s(D) P_c(H_0) \Bigg )\,,  
\label{eq:approx_post2}
\end{eqnarray} 
where 
\begin{eqnarray}
n_s(D, H_0) &\propto& \frac{1}{\sum_i W_i } \nonumber \times \sum_{i,~\Omega_i \in \Delta \Omega} \frac{W_i}{\sigma_{D_i}} \exp\left [-\frac{\left(D-D_L(z_i,H_0)\right)^2}{2\sigma_{D_i}^2} \right]\,.\nonumber\\
\label{eq:nd}
\end{eqnarray} 
Above, $\Delta \Omega$ represents a patch in the sky that can
vary in location and area for different GW
sources~\cite{Chen:2017wpg}, and $W_i \equiv W(D_i)$ is the aforementioned weight function that determines the likelihood of the $i^{th}$
galaxy to be the host. 

In the present work, we do not exhaust accounting for the various potential
astrophysical effects on $W_i$. For example, we assume that all
galaxies, regardless of their size, luminosity and type, are equally
likely hosts of BBHs. Although this is not a realistic assumption, it 
provides a simple framework to describe our method. We, however, 
account for the detector characteristics of the aLIGO-AdV
three-detector network and calculate the fraction of GW sources 
(10 $M_{\odot}$ + 10 $M_{\odot}$ BBH mergers) that will be detected 
in different sky patches and at varying depths. We average over the BBH
orbital orientations in space in order to obtain this detection fraction.
The resulting $W(D_i)$ is plotted as a function of distance 
for one of the sky patches in Fig.~\ref{weight_gal}.


Selection effects will indeed affect the above posterior by
influencing the integrand through the density of detected BBHs at varying
depths. The way we simulate this is detailed below. Its imprint on the
posterior appears through the normalization, as was shown in a general
setting studied in Ref.~\cite{mandel_weight}.


The spectroscopic redshifts of galaxies are obtained with high precision. Consequently, the derived luminosity distance errors from galaxy catalog are much smaller compared to the distance errors expected from GW measurements. In the limiting case where the error-bars are very small, it is clear that using the actual error-bars on the distance would result in a very fluctuating distribution. The two point-set distributions, obtained from the galaxies and the GW sources,  are best compared in a
coarse-grained manner. Therefore, instead of using the inferred
distance error for galaxies, we use a $\sigma_{Di}/D_L(z_i,H_0) =$ 10\% for
all galaxies, consistent with what we stated
earlier. However, the distance error for GW sources in our
simulations, $\sigma_{D}/D_0$ is taken to be their respective
measurement error, as explained below. 

For our calculations, $P_c(H_0)$, the prior probability for the Hubble parameter, is assumed to be uniform between 40-100 km/sec/Mpc. 
The sources are combined as before using the recursion for $\{{\cal S}_{\rm new}\} = \{{\cal S}, {\cal S}_{\rm old}\}$ using
\begin{eqnarray}
\nonumber
P(H_0\,|\,{\cal S}_{\rm new}) \propto \int dD && \Bigg( \exp \left [ -\frac{\displaystyle(D-D_0)^2}{\displaystyle 2\sigma_D^2} \right ]  \times \; n_s(D) P(H_0\,|\,{\cal S}_{\rm old}  ) \Bigg )\,.  
\label{eq:approx_post3}
\end{eqnarray} 
The upper limit on the integral is informed from the mock GW sample generated. For most patches $D_{\rm max}$ is chosen to be $\sim$3500 Mpc for the aLIGO-AdV network and $\sim$4500 Mpc for ET network.
Note that $n_s(D)$ comprises different sets of galaxies for different GW sources since these sources will, in general, be in different directions. To summarize our method we list below the steps one can follow to obtain the posterior distribution for $H_0$:
\begin{enumerate}
	\item Obtain the probability distribution over $D$ from GW measurements.  Note that, as mentioned earlier, in the realistic case one would obtain a distribution over both $\Omega$ and $D$, but for simplicity we use the distribution as given in Eq. \eqref{eq:likelihoodexample}.
	\item For each GW observation, obtain the galaxy distribution
	$n_s(D)$ from a sky-patch taken from the SDSS catalog that
	has support in the BBH distribution obtained in step 1
	(e.g., a sky patch in the direction of the BBH source with
	area similar to $\Delta \Omega$). 
	
	In our simulations, the sky localization error has the spread 
	mentioned in \S.~\ref{sec:intro} ($\Delta \Omega \in [3, 30]$~sq. deg.). However, whenever this error is less than 
	3 sq. deg., one can reset $\Delta \Omega$ to be equal to 3 sq. deg. and 
	retain those BBHs for further analysis. BBHs with errors greater than 30 sq. deg. can be dropped.
	Note that as stated earlier, we cite these number because we have confirmed the robustness of our method for these localisation areas. They are not claimed here to be `optimal' values of the sky patch area.
	\item Assume a prior for $H_0$. This could be either a uniform
	prior over some allowed range of admissible $H_0$ values or
	can be an informed prior coming from some other cosmological
	observations (e.g. Planck or HST measurements).
	\item The weighting function $W(D_i)$ used in this work
	is purely dependent on GW detection efficiency, and does not 
	depend on the galaxy distribution in any way. For every sky-patch of
	interest, we populate it uniformly on that section of the sphere with
	10-10 $M_\odot$ BBHs, such that there are 1600 sources in every
	distance bin. This number does not correspond to any realistic
	distribution of BBHs but is chosen merely to ensure that the error in
	detection probability estimated at each distance bin is less than 3\%
	(absolute). Additionally, we allow the cosine of the inclination angle
	to vary over 100 uniformly spaced values for every source. The SNR of
	each source is computed at each detector and only when it is above 8
	in at least two of them and the network SNR is 10 or higher do we classify it as detected. This is how we compute
	the detection probability in distance bins of 10 Mpc, in the range
	$D\in [50,4500]$ Mpc. Note that it varies from one sky patch to another
	because the antenna-patterns of the detectors (and even the network
	antenna pattern~\cite{Pai:2000zt}) vary across the sky. A geometric
	explanation for this weighting function is given in
	Ref.~\cite{Ghosh:2013yda}. (Additionally, we compute the network SNR
	of each source, and the average network SNR in each distance bin,
	which is used in the distance error estimation described below.)
	\item Plug these three distributions and the weighting function in Eq. \eqref{eq:approx_post2} to obtain the posterior over $H_0$.
	\item Combine the posterior distribution obtained from all the GW measurements to obtain the final $P(H_0|{\cal S})$.
\end{enumerate}
Note that in general one can obtain the posterior over any number of
cosmological parameters. In this work we have assumed a simple flat
LCDM model with two free parameters: $H_0$ and $\Omega_m$, and we fix
$\Omega_m=0.3$. In a future work we will also consider  $\Omega_m$ and
possibly the dark energy equation of state. Since we do not have
enough GW observations to test our method, we work with a simulated GW
catalog as also mentioned earlier. In the next subsection we discuss
how we simulate this GW catalog and also outline the key steps to
obtain $P(H_0|{\cal S})$ in this case.

\subsection{GW catalog: Second generation detectors}

We now describe the  method adopted for producing realistic catalogs of GW sources for near-future detector configurations. As mentioned earlier, we assume that the BBH sources are associated with galaxies. To get the galaxy samples we chose different
sky patches from the SDSS database. Our sky patches were obtained from a conical search (limited in
redshift and solid-angle), in the  SDSS catalog. In principle one can choose from a variety of shapes for the sky patch,
but in this first study we chose this simple shape to focus more on
the main idea behind the method.  

The first step to construct the GW catalog is to obtain the weight function $W(D)$. This was outlined in the previous section. 
The detector characteristics (e.g. the noise sensitivity), the distance to the source and its sky position, and the orientation of the
plane of the binary with respect to the detector, all play an important role in determining
whether the event will be detected by the GW detectors  and hence in determining $W(D)$ (see, e.g., Refs.~\cite{Pai:2000zt,Bose:2011} and the references therein). 
Note that in this work we have studied a particular population of BBH sources, but the analysis can be extended to other kinds of binary sources, such as NS-BH or NS-NS systems; moreover, BBH with somewhat higher masses will provide smaller error at any given redshift and more redshift depth. 

\begin{figure}[h]
	\centering  
	{\includegraphics[width=0.5\textwidth]{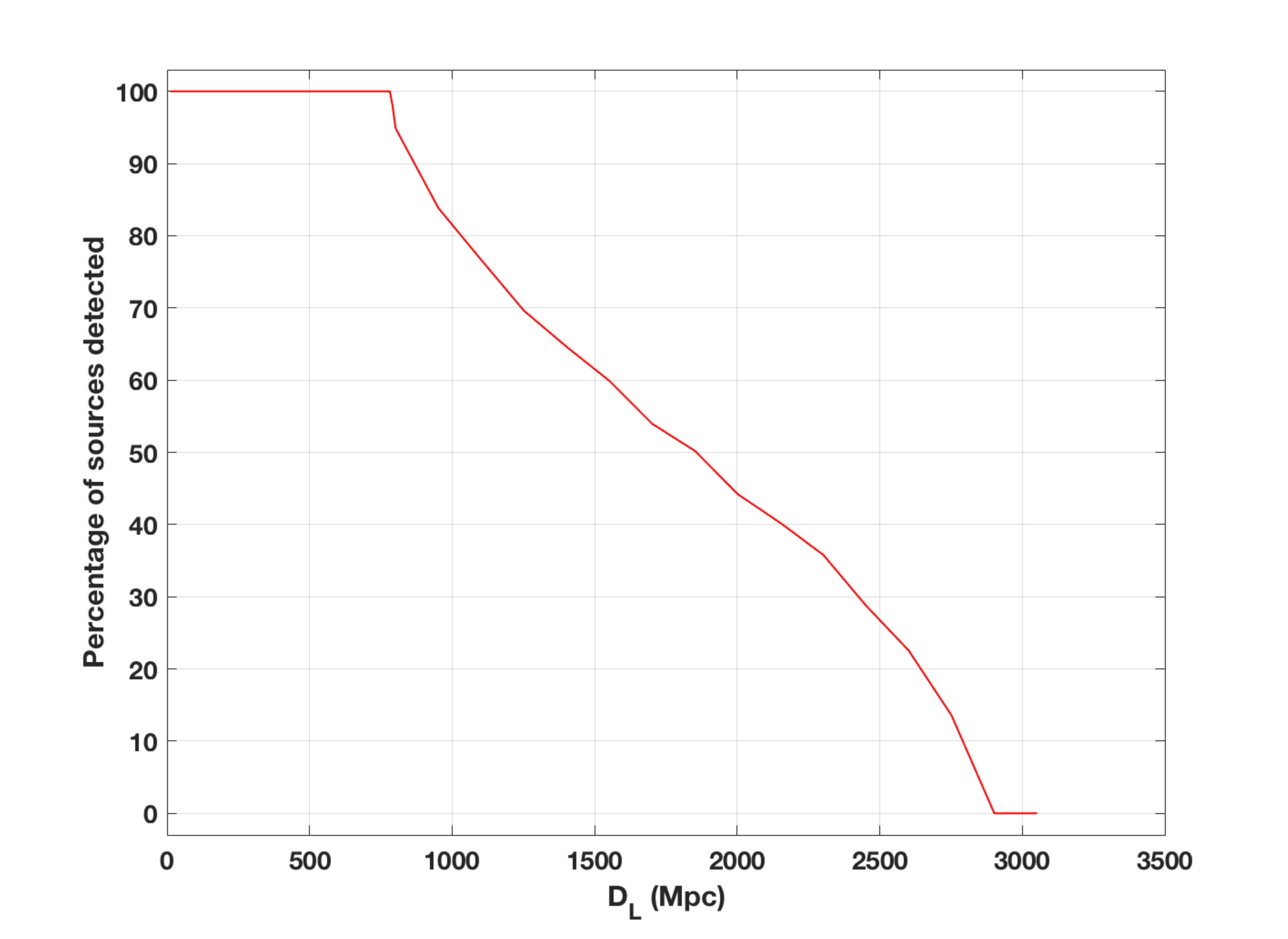}}
	\caption{The weights $W(D_i)$ assigned to the galaxies in a single sky patch is plotted as a function of
		distance for one of the sky patches for the aLIGO-AdV network.}
	\label{weight_gal}
\end{figure}
%

\begin{figure*}[!htbp]
	\includegraphics[height=0.30\textwidth]{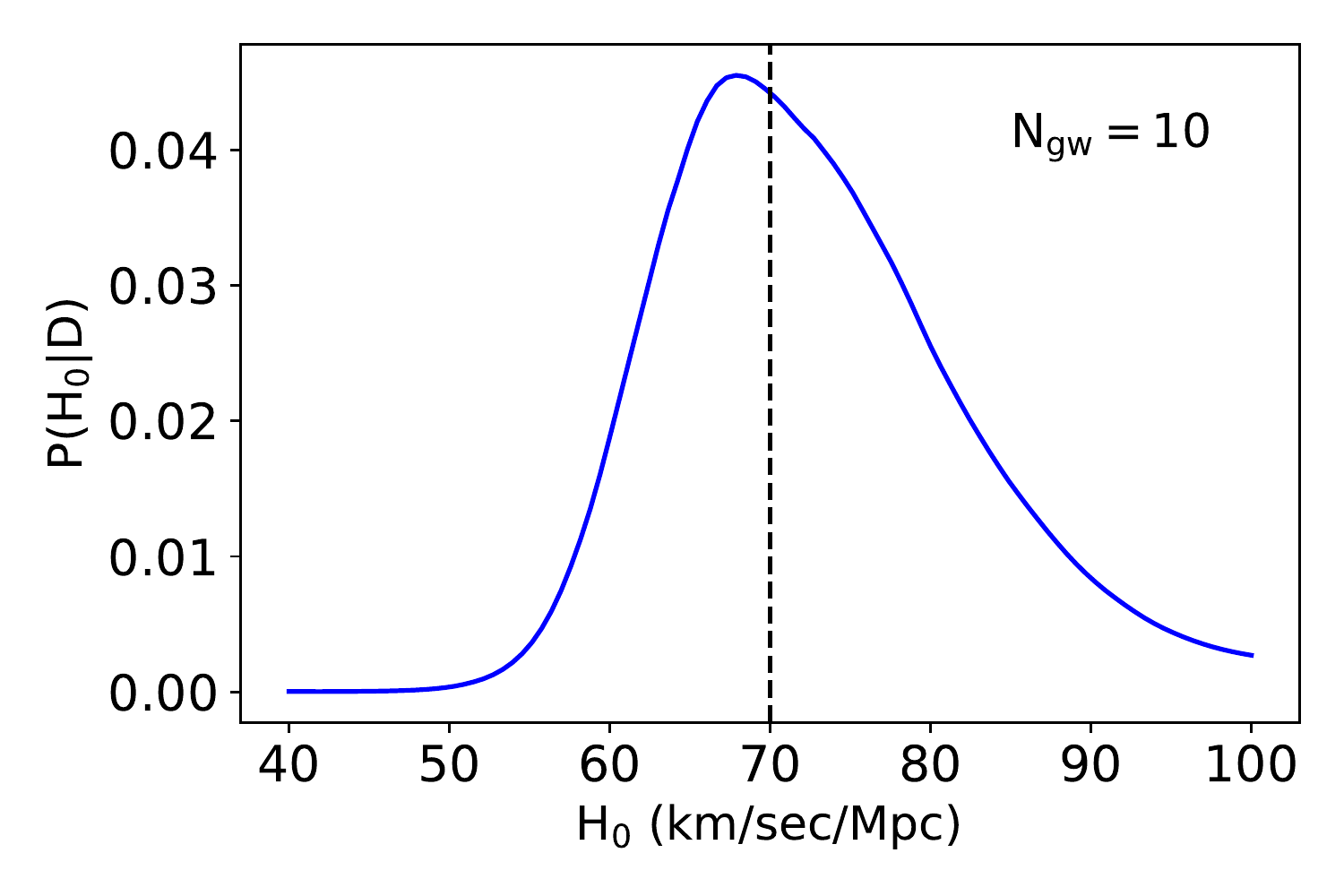}
	\includegraphics[height=0.30\textwidth]{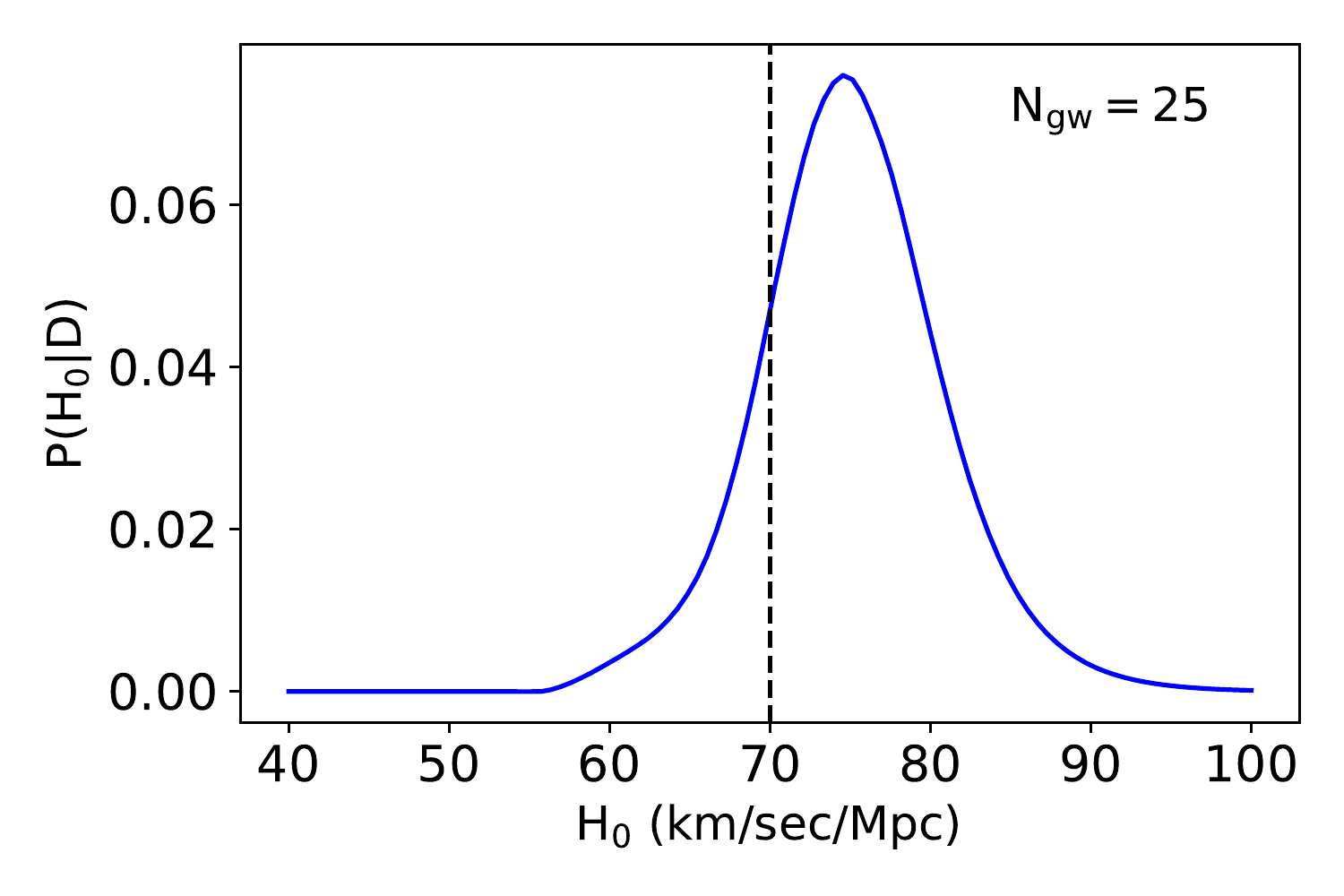}\\
	\includegraphics[height=0.30\textwidth]{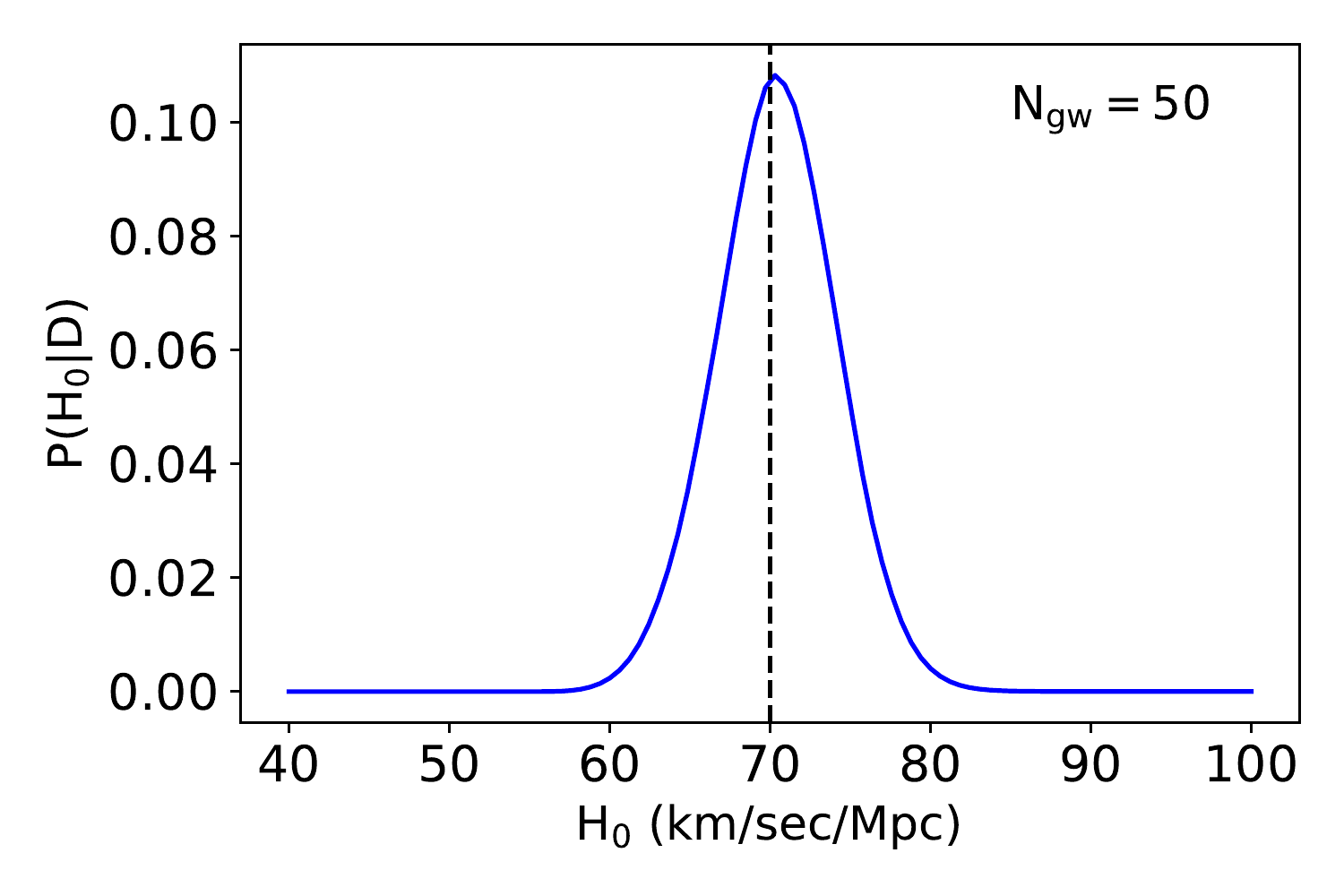}
	\includegraphics[height=0.30\textwidth]{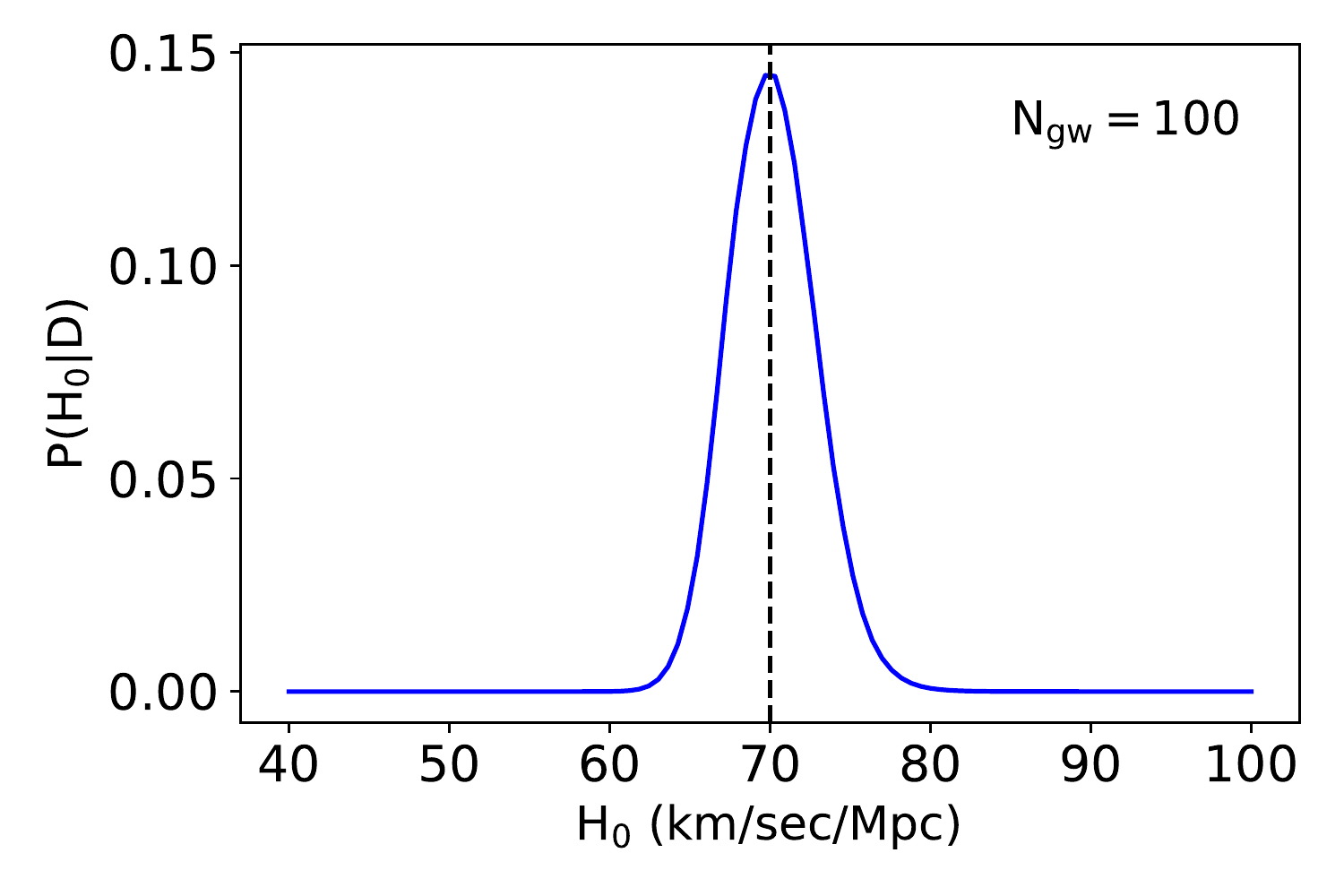}
	\caption{The plots show the projected normalized posterior probability distribution for $H_0$ obtained from the aLIGO-AdV network. Dashed line at 70 km/sec/Mpc, represents the value of $H_0$ used for simulation (fiducial input). The number of GW sources used, is mentioned in each panel. The value of $H_0$ at the peak of the posterior distribution is (clockwise from upper-left panel) 67.8 km/sec/Mpc, 74.6 km/sec/Mpc, 70.3 km/sec/Mpc and 70.0 km/sec/Mpc respectively. For all plots the sky patches have localisation error $\Delta \Omega \in [3, 30]$~sq. deg..
		\label{aLIGO}}
\end{figure*}

To obtain the weight function $W(D)$ mentioned above, we evaluate
what fraction of our BBH population (which is randomly oriented in space to create a realistic population) will have SNRs above a threshold value in our detector-network of interest. We
set this threshold such that the SNR is at least 8 in two of the three detectors
and the network SNR is at least 10. 
This criterion is used to estimate what fraction of BBH sources at some distance
will be detectable in GWs. We take the distance error to be 30\%
for a BBH at a network SNR of 20~\cite{local2,aghosh,nissanke}. For a BBH signal
with network SNR = $\rho$, its percentage distance error scales roughly as
30\%$\times (20/\rho)$. For ET, the fraction of BBHs detected, remains 100\% to a greater distance than
that corresponding to the second generation aLIGO-AdV network. Moreover, the SNR of the same BBH is about 10 times higher, and
distance error about 10 times lower, in ET than in aLIGO-AdV.
Once we have the fraction of BBH sources that are 
detectable by the GW network as a function of distance, we use it
to obtain $P(H_0|{\cal S})$ as outlined below:
\begin{enumerate}
	
	\item 
	In the sky patch of interest, say we have $N$ galaxies distributed over some redshifts.
	
	\item 
	Map all the redshifts to distances using Eq. \eqref{dl} where $H_0$ is now a free parameter ($\Omega_m=0.3$). We assign each galaxy a weight based on what fraction of BBH sources to its distance would be detectable by the GW detector network (see Fig.~\ref{weight_gal}). Next we construct the distance distribution for the galaxies as in Eq. \eqref{eq:nd}. 
	
	\item 
	Assume a prior distribution over $H_0$. Here we use a uniform prior over 40-100 km/sec/Mpc.
	
	\item
	Now, we go back to the galaxy redshift catalog and select a galaxy randomly, so any of the $N$ galaxies is equally
	likely to be picked.
	
	\item
	Say, the chosen galaxy falls in the $j^{\rm th}$ bin. We put a BBH source in this galaxy with
	a probability $f_j$, where $f_j$ is the fraction of detectable sources at that distance
	(this is done by throwing a uniform random number $q$ between $[0,1]$
	and putting a source in the galaxy if $f_j>q$). 
	
	\item
	Once we have the BBH source, we first map its redshift to a `true' distance $D_m$ (evaluated from Eq. \eqref{dl}) by assuming $\Omega_m = 0.3$ and $H_0 = 70$ km/sec/Mpc. We also know what is the expected SNR of this BBH merger and we translate that to an error ($\sigma_{D}$) on the distance as discussed earlier in this subsection.
	
	\item 
	We randomly sample from a Gaussian distribution, centered at $D_m$ and with standard deviation equal to the distance error, both obtained in the previous step.
	
	\item 
	Once we have the `observed' $D_0$ and the corresponding error we construct the probability distribution for this `measurement' (Eq. \eqref{eq:likelihoodexample}). 
	
	\item Now that we have all the required probability distributions we plug them in Eq. \eqref{eq:approx_post2} to obtain the posterior over $H_0$.
	
\end{enumerate}
Note that, as mentioned earlier, the weights assigned to galaxies depend on the direction of the sky patch (as GW detectors do not have isotropic efficiency).
The weights given to the galaxies, for the three-detector aLIGO-AdV network, as a function of redshift for one of the sky patches
is shown in Fig.~\ref{weight_gal}. In the figure one can see that the weight drops substantially beyond a distance of $\sim$2000~Mpc.
This drop occurs because of the SNR threshold we set on the GW events~\cite{Ghosh:2013yda}.
\begin{table}
	\begin{center}
		\resizebox{0.30\columnwidth}{20.5mm}{%
			\begin{tabular}{ c c c } 
				\hline
				$N_{\rm GW}$ & $H_0$ & $H_0$\\
				$ $ & {\small (aLIGO-AdV)} & {\small (ET)}\\
				
				\hline
				\hline
				\\[0.05mm]
				$10$ & $67.8^{+12.0}_{-6.1}$  & $69.1^{+8.2}_{-9.9}$\\[1.5mm]
				$25$ & $74.6^{+5.7}_{-5.4}$ & $75.3^{+4.9}_{-5.1}$\\[1.5mm]
				$50$ & $70.3^{+4.1}_{-3.6}$ & $68.2^{+3.1}_{-2.9}$\\[1.5mm] 
				$100$ & $70.0^{+2.7}_{-2.8}$ & $70.1^{+2.6}_{-2.3}$\\[1.5mm] 
				\hline
				\hline
			\end{tabular}
		}
		\caption{The table shows the estimated values of the Hubble constant in units of
			km/sec/Mpc, from the aLIGO-AdV network (second column), and the ET network (third column) along with error-bars obtained by considering the threshold value for $P(H_0 \,|\, {\cal S})$ that encloses  68\% probability region around the peak of the distribution. Different rows show
			different number of GW sources. The binaries are simulated to have random orientations, and the distance errors are determined by averaging over the orientations.}
		\label{tab2}
	\end{center}
\end{table}

\subsection{GW catalog: Third generation detector}

We next apply the same method to a third generation detector like the Einstein Telescope (ET) \cite{et}
to see how these estimates will change if the
same BBH sources are obtained with much higher SNRs. We assume a
triangular configuration for ET as discussed in \cite{broeckET} with
three interferometers located at the same sites as LIGO-Hanford, 
LIGO-Livingston and Virgo, respectively.. The way the catalog is
generated is the same as in the previous section but instead of
aLIGO-AdV the Einstein Telescope is used as a GW detector-network. 
Unlike the second generation detectors, ET will observe the same BBH
sources with much smaller distance
errors. It will also detect sources to larger redshifts (in a future work, we plan to study ET BBH sources to analyze how well ET may be able to constrain the dark
energy equation of state~\cite{gw_em_cosmo}). 
A plot similar to Fig.~\ref{weight_gal} would show 100\% sources recovered to distances of $\sim$ 5000 Mpc. 

\section{Results}

Once we know how to populate the GW catalog using the SDSS galaxies as hosts we generate multiple such catalogs with varying number of GW sources. These sources are then assigned distances and distance errors, and the galaxies are also weighted to account for the detection efficiency of the GW detectors, using the method outlined in the sections \S~\ref{sec:method} and \S~\ref{sec:num_sim}.  Once we have these distributions we estimate the posterior distribution of the cosmological parameters given the (mock-)GW data (Eq.~\eqref{eq:approx_post2}). 
%

\begin{figure*}[!htbp]
	\includegraphics[height=0.30\textwidth]{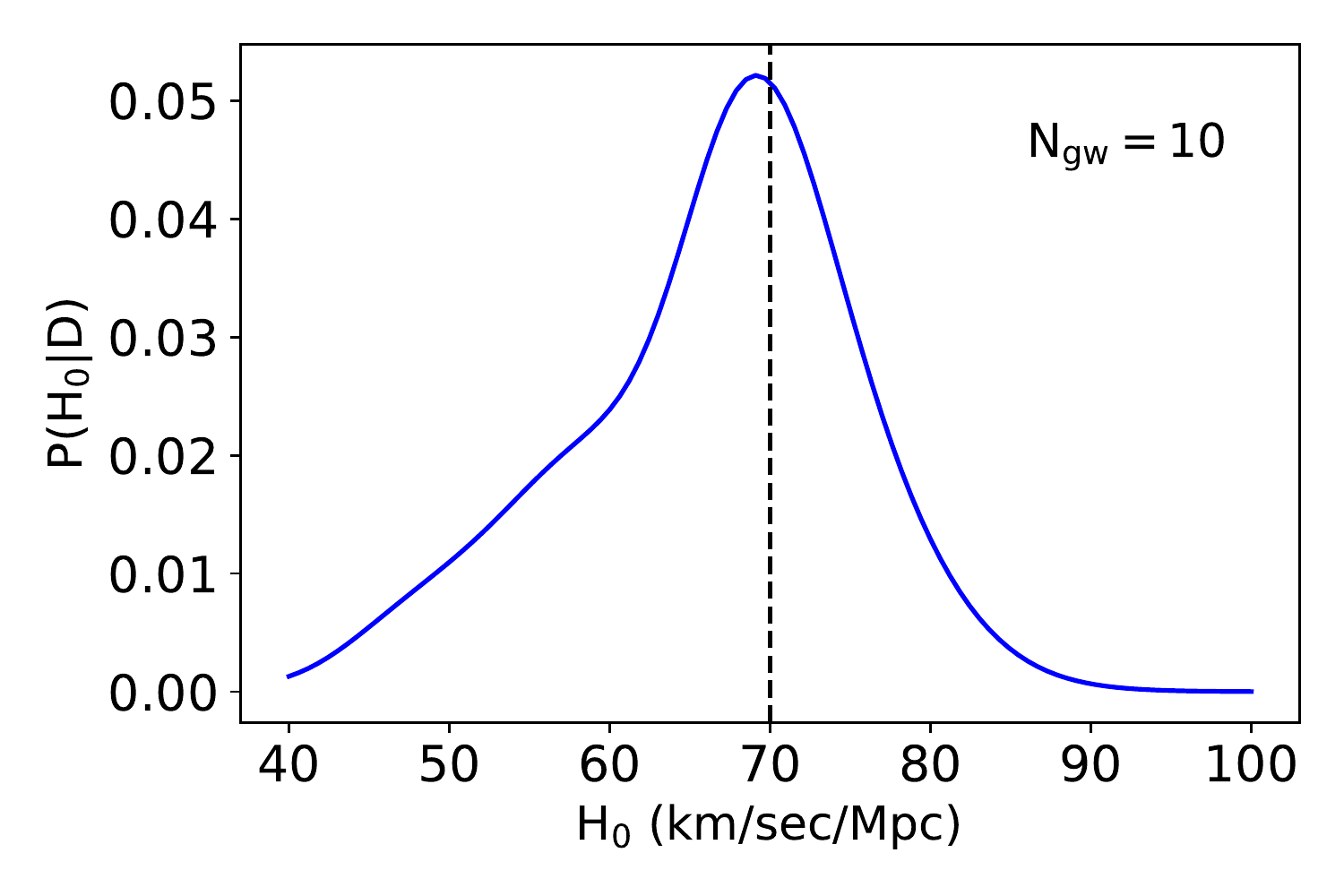}
	\includegraphics[height=0.30\textwidth]{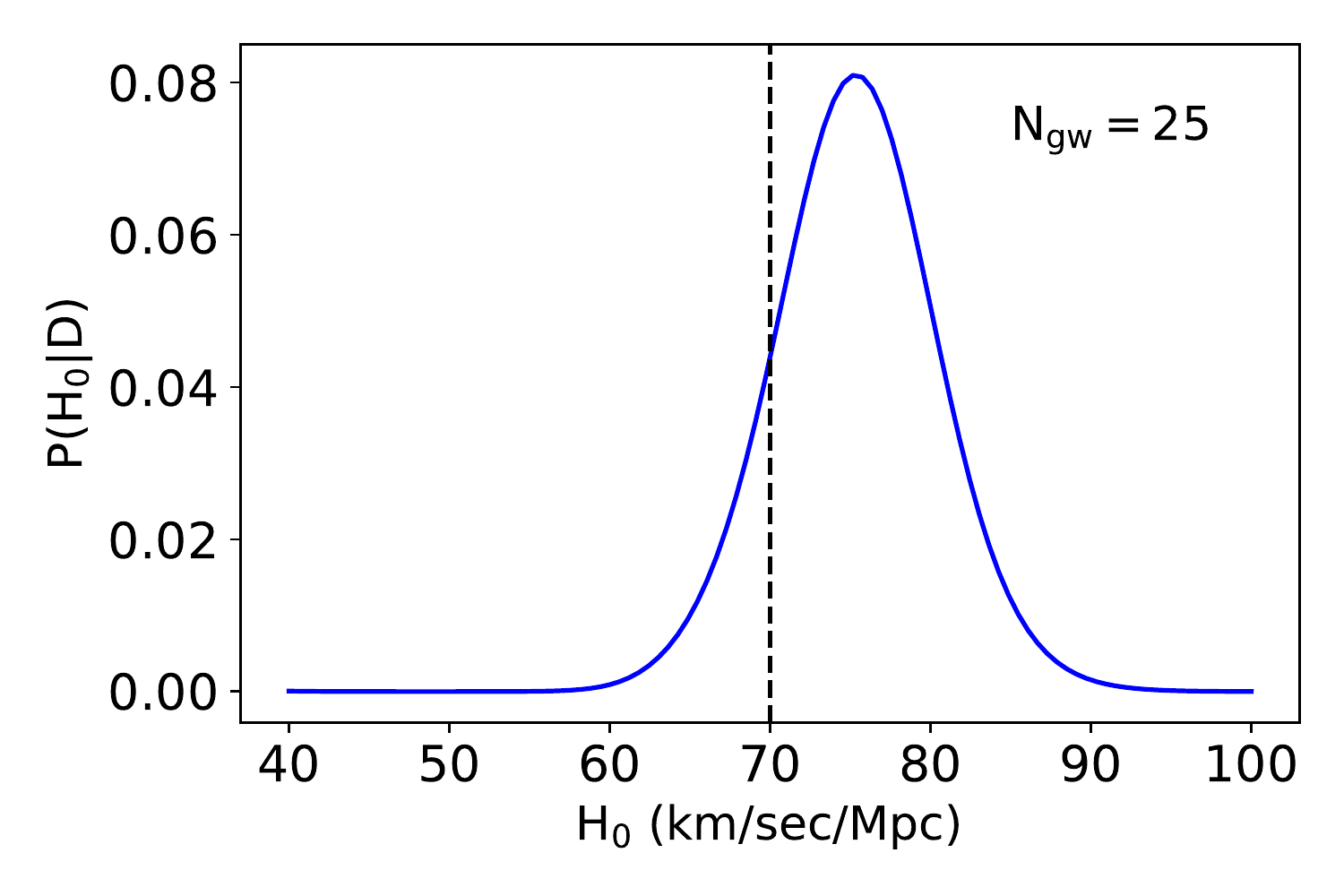}\\
	\includegraphics[height=0.30\textwidth]{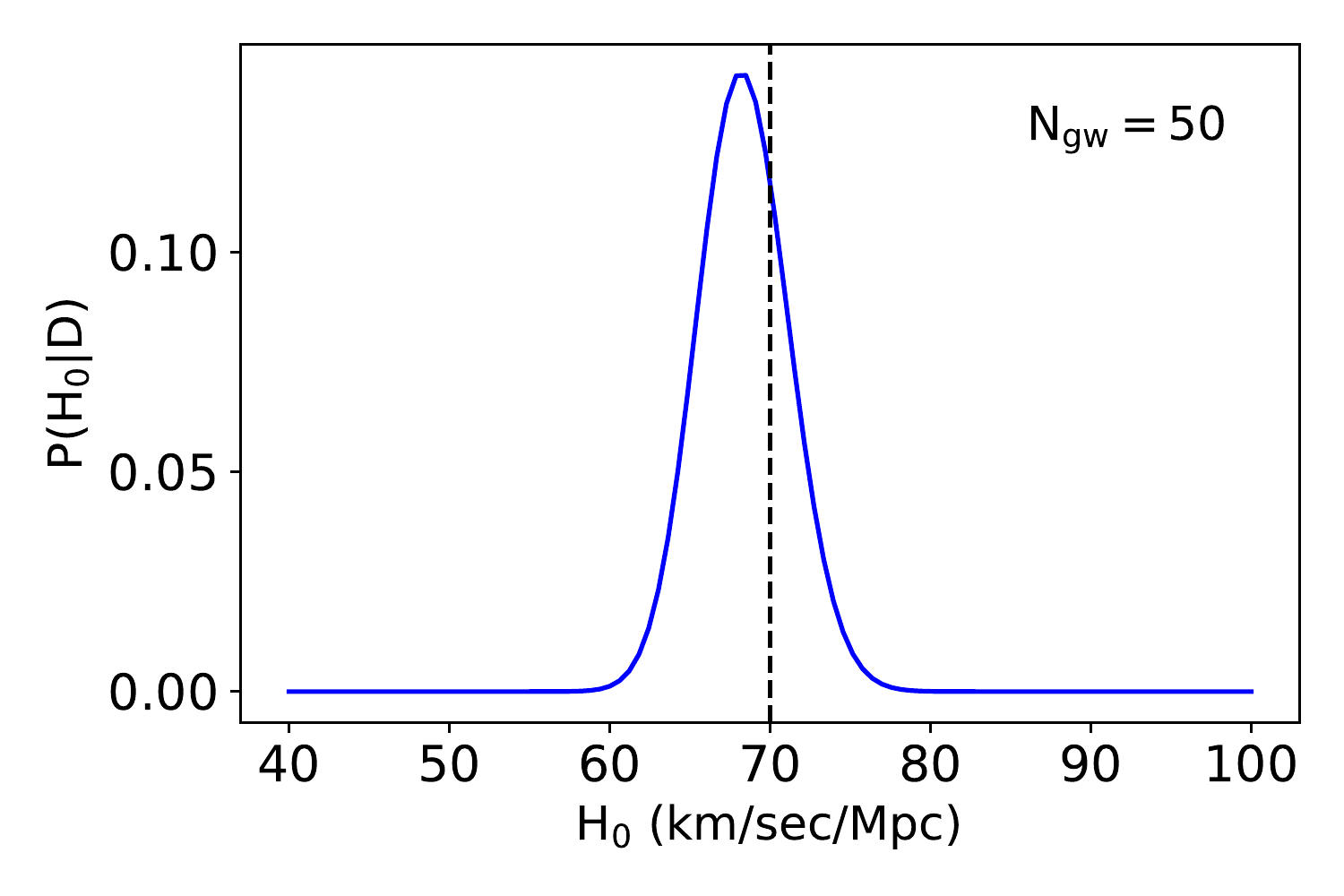}
	\includegraphics[height=0.30\textwidth]{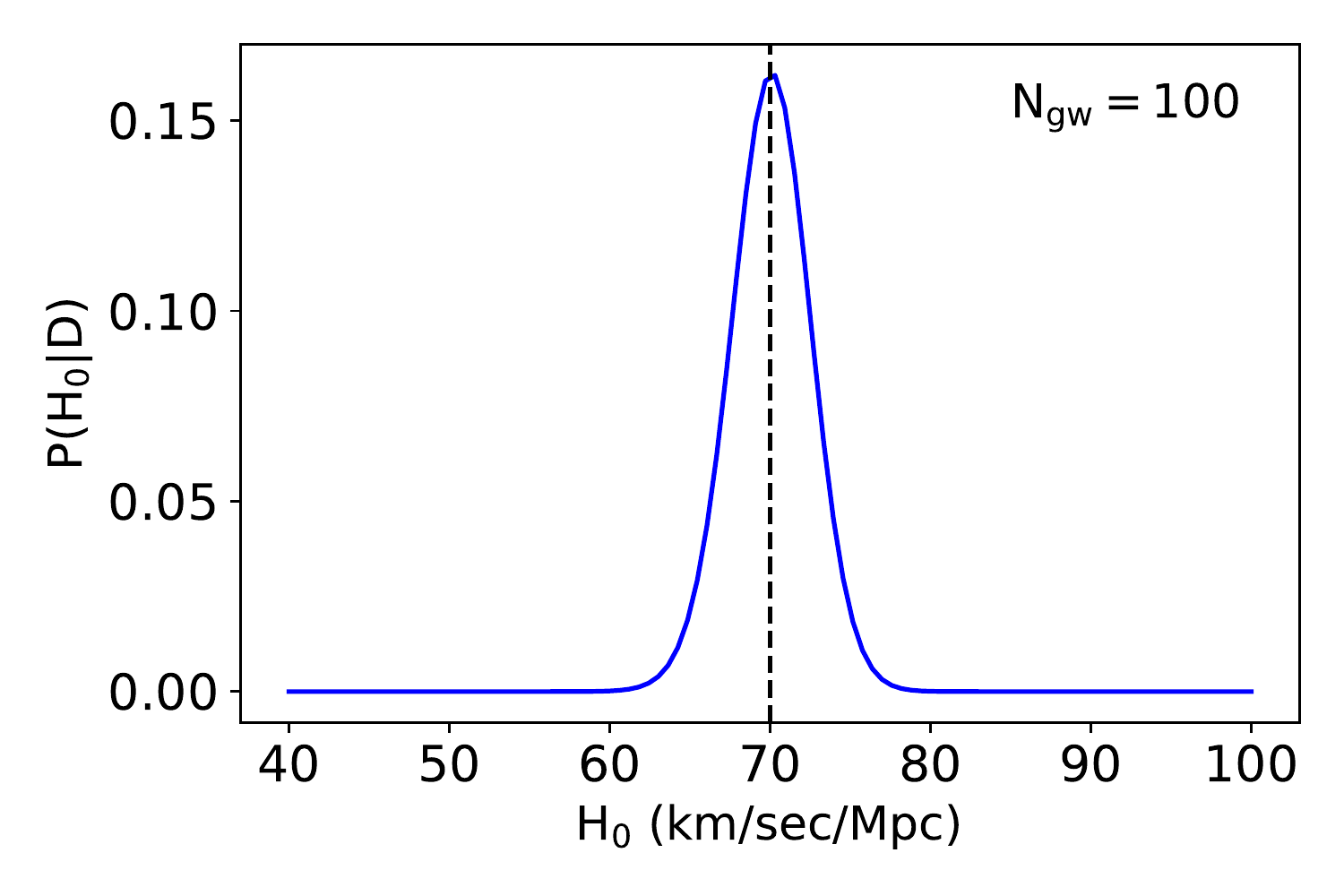}
	\caption{The plots show the normalized posterior probability distribution for $H_0$ obtained from the third generation detector ET . Dashed line at 70 km/sec/Mpc, represents the value of $H_0$ used for simulation (fiducial input). The number of GW sources used, is mentioned in each panel. The value of $H_0$ at the peak of the posterior distribution is (clockwise from upper-left panel) 69.1 km/sec/Mpc, 75.3 km/sec/Mpc, 68.2 km/sec/Mpc and 70.1 km/sec/Mpc respectively. For all plots the sky patches have localisation error $\Delta \Omega \in [3, 30]$~sq. deg..
		\label{ET}}
\end{figure*}

We perform the analysis for aLIGO-AdV network as well as for a third generation ET network. Note that the estimates and plots shown are form a particular set of sky patches and a specific realization of the GW catalog. The estimate and the posterior distribution of $H_0$ will be different for a different setting. The results will converge for large number of GW sources. 

Results obtained for the second generation detector network are shown in Table~\ref{tab2} and Fig. \ref{aLIGO}. We estimate the posterior distribution for $H_0$ using varying number of GW sources and we report the $H_0$ value that corresponds to the peak of the posterior distribution. We also quote error-bars obtained by considering the threshold value for $P(H_0 \,|\, {\cal S})$ that encloses  68\% probability region around the peak of the distribution. With the aLIGO-AdV network we find that one can constrain $H_0$ with
an accuracy of $\sim$ 8\% with as few as 25 GW sources. 
The integral in Eq. \eqref{eq:approx_post2} would peak for the value
of $H_0$ for which the GW probability distribution in $D$ (Eq.~(\ref{eq:likelihoodexample})) has
maximum overlap with the probability distribution in $D$ obtained from
galaxies (Eq. \eqref{eq:nd}). Note that when there are multiple regions of
comparable clustering in the galaxy distribution, the posterior in
$H_0$ can be multi-modal. But this problem can be addressed by combining
multiple GW measurements. Combining measurements is similar to
stacking histograms and we expect that eventually, a peak emerges
around the correct value of $H_0$.

Results obtained by repeating the analysis for a third generation detector (ET) network are also shown in Table~\ref{tab2} and in Fig. \ref{ET}. In this case we find that $H_0$ is constrained to an accuracy of about 7\% with 25
GW sources. Note that in this case since we can go much deeper in redshift, we can also get estimates of parameters like $\Omega_m$ (which we have fixed to 0.3 here), or the equation of state for dark energy (in the case of non-LCDM models). But at higher redshifts, the incompleteness of the galaxy catalog may potentially be a problem and one would have to address this in the formulation more carefully. We leave this exercise for  future work. 
\section{Discussion}
GW signals from coalescing binaries will provide distance measurements
that are complementary to the electro-magnetic standard candle measurements
used to constrain cosmological parameters. Additionally these
measurements do not suffer from the calibration error that is one of the major systematic uncertainties that plagues the supernovae
measurements. (Note that GW detectors are affected by intrinsic
calibration uncertainties that are at present no more than 10\% in strain
amplitude and 10$^\circ$ in phase~\cite{gw150914properties}.
Efforts are on, however, to reduce these errors~\cite{gw150914calibration}.) But doing cosmology with inspiraling binaries requires the use of data obtained from EM surveys
since GW measurements from these binaries alone can not provide redshift information of the source, which is imperative
to constrain the distance-redshift relation. Some of these GW sources (like NS-NS mergers)
are expected to have EM counterparts,
and a coordinated EM observation of the source can obtain redshifts for these events, as was done for the NS-NS event observed by the LIGO-VIRGO network. But this
method will not work for mergers that are not accompanied by an electromagnetic event, for
electromagnetic events that are too short-lived, or for very far off sources. Many methods have been suggested in the past
to address this problem, as discussed in the introduction. 

In this work we proposed to use the spatial clustering of galaxies, as seen through many large-scale surveys, to infer the spatial clustering of GW sources. We have introduced a general Bayesian formulation for extracting cosmological information using GW observation and galaxy clustering data. The general formulation takes into account the possibility of complex constraints on spatial location of GW sources, which includes angular location as well as distance information. The formulation was then simplified to ascertain the efficacy of this technique by generating simulated GW data. This mock GW catalog was generated using the galaxies from the SDSS survey as `host' galaxies. For this work we considered only 10 $M_{\odot}$ + 10 $M_{\odot}$ BH-BH binaries. The GW sources were are then analyzed with the SDSS galaxies as the hosts of GW events. 

We estimated the posterior distribution of the Hubble parameter $H_0$ by analyzing the expected GW measurements from second generation detector network aLIGO-AdV, where we included the information about
the orientation of the binary, sky position, detector characteristics, etc., and we found that one can constrain $H_0$ with an accuracy of $\sim$ 8\% with just 25 sources (Fig. \ref{aLIGO}). Third generation detectors like the Einstein Telescope will see the same sources with much higher SNR and, therefore, smaller distance errors. It will also detect sources to larger redshifts. Here,
however, we restricted ET observations to BBH sources up to a similar
depth, specifically, with $z \leq 0.6$ and we showed that measurements
from the third generation detector will be able to constrain $H_0$ to
an accuracy of $\sim$ 7\% with 25 detections (Fig. \ref{ET}). 

Hence we have shown that it should be possible to obtain excellent constraints on the Hubble parameter with the near-future GW detector configurations. However, the same technique could be used for future experiments to extract the matter density parameter ($\Omega_m$) or properties of dark energy, like the equation of state etc. This would be possible when the data acquires more precision and sufficient redshift depth.

While working on this
paper we came across Ref.~~\cite{oguri}, which is somewhat similar in
spirit to this work in that the author sets constraints on cosmological parameters by using the cross-correlation between observations from ET and the Euclid survey. We would like to note that in addition to giving constraints for the current second generation ground-based detectors, our treatment is more realistic since we also take into account the detector characteristics in detail.

Note that we have made many simplifying assumptions in this work. We have
assumed that the spatial distribution of our GW sources (BBH) is
identical to that of galaxies. In reality, the merger rates of
coalescing binaries may depend in some hitherto unknown manner on the
source galaxies. For example, the number of GW sources in a galaxy
should scale with the number of stars in a galaxy, so our assumption
that each galaxy contributes equally to the galaxy distribution
function may not translate to it contributing equally to the GW source
function. This can be taken into account if we utilize information about
the luminosity of a galaxy, which we can use while assigning how
much it contributes to the galaxy source function. However, it is likely that the GW rates could
also depend on the galaxy type, i.e., on whether the galaxy is spiral
or elliptical. Since spiral and elliptical galaxies cluster
differently, this would have an impact on parameter
estimation. Considering that any sufficiently prominent peak of galaxy
cluster would roughly contain an equal mixture of different types of
galaxies, and the fact that our method extracts maximum information
from the dominant peaks in the galaxy distribution, to zeroth order
our method should work fine. However, to obtain more detailed
information (such as parameters of dark energy) from this method would
require addressing these issues in some detail. This is beyond the
scope of this investigation and will be followed up in a future work.

Furthermore an addition that may become important
in the future for an analysis like this is the information from population synthesis models. These models
predict the source
mass distribution for NS-NS, NS-BH or BBH binaries. They also predict merger rates for these
binaries. But since most of these models are degenerate and a large number of GW  detections are
required to narrow in on some preferred model, we have not included them in the current analysis.   
A further important improvement may come from all-sky galaxy
catalogs. Since GW detectors are
not directional, a coalescing binary can be observed in most parts of the
sky. This is more true for GW detector networks. Even so, it may happen that some sources fall in
sky areas that are not covered by galaxy surveys yet. In such a case we will not be able to use such 
GW signals in our analysis.  These and other issues will be addressed in a future work. 

\begin{acknowledgements}
	We thank the anonymous referee for suggesting useful changes to the
	manuscript which have improved our analysis significantly. We would
	also like to thank Varun Sahni, Sheelu Abraham, Timothee Delubac, 
	Takahiro Tanaka and John Veitch for discussions at various stages 
	of this project. We thank Will Farr and Archisman
	Ghosh for carefully reading the manuscript and making several useful
	suggestions. RN acknowledges the support of the Japan Society for the Promotion of Science (JSPS) fellowship, Grant No. 16F16025. RN also thanks the Inter University Centre for Astronomy and Astrophysics (Pune) for hospitality, where part of this work was done. This work is supported in part by NSF grants PHY-1206108 and PHY-1506497, and the Navajbai Ratan Tata Trust. We made use of the publicly available data from SDSS. Funding for the SDSS and SDSS-II has been provided by the Alfred P. Sloan Foundation, the Participating Institutions, the National Science Foundation, the U.S. Department of Energy, the National Aeronautics and Space Administration, the Japanese Monbukagakusho, the Max Planck Society, and the Higher Education Funding Council for England. The SDSS Web Site is http://www.sdss.org/.
	The SDSS is managed by the Astrophysical Research Consortium (ARC) for the Participating Institutions. The Participating Institutions are The University of Chicago, Fermilab, the Institute for Advanced Study, the Japan Participation Group, The Johns Hopkins University, Los Alamos National Laboratory, the Max-Planck-Institute for Astronomy (MPIA), the Max-Planck-Institute for Astrophysics (MPA), New Mexico State University, University of Pittsburgh, Princeton University, the United States Naval Observatory, and the University of Washington.
\end{acknowledgements}


\begin{thebibliography}{99}
	\bibitem{de} B. Ratra \& P. J. E. Peebles, \prd ~{\bf 37}, 3406 (1988);
	C. Wetterich, Nucl. Phys. B ~{\bf 302}, 668 (1988);
	J. Ellis, S. Kalara, K.A. Olive \& C. Wetterich, \plb ~{\bf 228}, 264 (1989);
	R. R. Caldwell, R. Dave \& P. J. Steinhardt, \prl ~{\bf 80}, 1582 (1998);
	V. Sahni \& A. Starobinsky, \ijmpd ~{\bf 15}, 2105 (2006);
	E. J. Copeland, M. Sami \& S. Tsujikawa, \ijmpd ~{\bf 15}, 1753 (2006);
	J. Frieman M.Turner \& D. Huterer \ARAnA ~{\bf 46}, 385 (2008);
	R. R. Caldwell \& M. Kamionkowski, \ARNPs ~{\bf 59}, 397 (2009;
	T. Clifton, P. G. Ferreira, A. Padilla \& C. Skordis \pr ~{\bf 513}, 1 (2012).
	
	\bibitem{local2} P. Ajith \& Sukanta Bose, \prd ~{\bf 79}, 084032 (2009).
	
	\bibitem{aghosh} A. Ghosh, P. Ajith \& W. del Pozzo \prd ~{\bf 94}, 104070 (2016).
	
	\bibitem{nissanke} S. Nissanke, D. E. Holz, S. A. Hughes, N. Dalal, \& J. L. Sievers \apj ~{\bf 725}, 496 (2010).
	
	\bibitem{ligo_H0} B. P. Abbott et al., Nature ~{\bf 551} 85 (2017).
	
	\bibitem{rev_wein} D. H. Weinberg, M. J. Mortonson, D. J. Eisenstein, C. Hirata, A. G. Riess, \& E. Rozo,
	\pr ~{\bf 530} (2), 87 (2013).
	
	\bibitem{de_obs} R. Amanullah et al., \apj ~{\bf 716}, 712 (2010);
	A. Lewis \& S. Bridle, \prd ~{\bf 66}, 103511 (2002);
	M. J. Mortonson, D. Huterer \& W. Hu, \prd ~{\bf 82} 063004 (2010);
	C. Blake et al., \mnras ~{\bf 418}, 1707 (2011);
	L. Anderson et al., \mnras ~{\bf 427}, 3435 (2012);
	U. Seljak, A. Makarov, R. Mandelbaum, C. M. Hirata, N. Padmanabhan, P. McDonald, M. R. Blanton, M. Tegmark, N. A. Bahcall,  J. Brinkmann, \prd ~{\bf 71}, 043511 (2005).
	
	\bibitem{sneia}  Supernova Search Team collaboration: A. G. Riess et al., 
	\aj ~{\bf 116}, 1009 (1998); Supernova Cosmology Project collaboration:
	S. Perlmutter et al., \apj ~{\bf 517}, 565 (1999).
	
	\bibitem{des} www.darkenergysurvey.org
	
	\bibitem{pans} www.pan-starrs.ifa.hawaii.edu
	
	\bibitem{lsst} www.lsst.org
	
	\bibitem{sneia_syst1} A. Conley et al., \apjs ~{\bf 192}, 1 (2011);
	D. Scolnic et al., \apj ~{\bf 795}, 45 (2014);
	J. Newling, B. Bassett, R. Hlozek, M. Kunz, M. Smith \& M. Varughese,  \mnras ~{\bf 421} (2), 913 (2012).
	
	\bibitem{sneia_syst2} 
	P. L. Kelly, M. Hicken, D. L. Burke, K. S. Mandel, R. B. Kirshner,  \apj ~{\bf 715}, no. 2 743 (2010).
	
	\bibitem{sneia_syst3}
	R. J. Foley et al., \aj ~{\bf 143}, no. 5 113 (2012).
	
	\bibitem{hst} 
	W. L. Freedman et al., \apj ~{\bf 553}, no. 1 47 (2001).
	
	\bibitem{planck14} P. A. R. Ade et al. (Planck collaboration), \AnA
	~{\bf 571} A1 (2014).
	
	\bibitem{riess11}  A. G. Riess, L. Macri, S. Casertano, H. Lampeitl, H. C. Ferguson, A. V. Filippenko, S. W. Jha, W. Li,  R. Chornock, 
	\apj ~{\bf 730}, 119 (2011)
	Erratum: \apj ~{\bf 732}, 129 (2011).
	
	\bibitem{planck16} Planck   Collaboration: N. Aghanim et al.,
	\AnA ~{\bf 596}, A107 (2016).
	
	\bibitem{riess16} A. G. Riess et al., \apj ~{\bf 826} 56 (2016).
	
	\bibitem{valentino} E. D. Valentino, A. Melchiorri \& J. Silk,
	\plb ~{\bf 761}, 242 (2016).
	
	\bibitem{sneia_bao} W. J. Percival, S. Cole, D. J. Eisenstein, R. C. Nichol, J. A. Peacock, A. C. Pope \& A. S. Szalay, \mnras ~{\bf 381}, 1053
	(2007); R. Lazkoz, S. Nesseris \& L. Perivolaropoulos, JCAP ~{\bf 07}
	012 (2008).
	
	\bibitem{gw150914} B. P. Abbott et al., \prl ~{\bf 116}, 061102 (2016); 
	
	\bibitem{gw150914properties} B. P. Abbott et al., \prl ~{\bf 116}, 241102 (2016); 
	
	\bibitem{gw150914calibration} B. P. Abbott et al., \prd ~{\bf 95}, 062003 (2017); 
	
	\bibitem{gw_det} LIGO Scientific Collaboration and Virgo Collaboration:
	B. P. Abbott et al., \prl ~{\bf 116}, 241103 (2016);
	B. P. Abbott et al., \prl ~{\bf 118}, 221101 (2017);
	B. P. Abbott et al., \prl ~{\bf 119}, 141101 (2017);
	B. P. Abbott et al., \prl ~{\bf 119}, 161101 (2017);
	B. P. Abbott et al., \apjl ~{\bf 851} L35 (2017).
	
	\bibitem{schutz} B. F. Schutz, Nature {\bf 323}, 310 (1986).
	
	\bibitem{ligo} http://ligo.org
	
	\bibitem{virgo} http://www.virgo-gw.eu
	
	\bibitem{local1} LIGO Scientific Collaboration and Virgo Collaboration:
	B. P. Abbott et al., \apj ~{\bf 826}, no.1, L13 (2016).
	
	\bibitem{gw_em} LIGO Scientific Collaboration and Virgo Collaboration:
	B. P. Abbott et al., \apjl ~{\bf 848} 2 (2017).
	
	\bibitem{kagra} KAGRA Collaboration: Yoichi Aso et al., \prd ~{\bf 88}, 043007  (2013).
	
	\bibitem{ligo-I} C. S. Unnikrishnan, \ijmpd ~{\bf 22}, 1341010 (2013).
	
	\bibitem{et} http://www.et-gw.eu
	
	\bibitem{sgrb_NS} N. Tanvir, A. J. Levan, A. S. Fruchter, J. Hjorth, R. A. Hounsell, K. Wiersema \& R. L. Tunnicliffe, 
	Nature ~{\bf 500}, 547 (2013);
	E. Berger, W. Fong \& R. Chornock, \apjl ~{\bf 774} L23 (2013).
	
	\bibitem{EM_gw} LIGO Scientific Collaboration, Virgo Collaboration, Fermi Gamma-Ray Burst Monitor, INTEGRAL, 
	~ \apjl L13 ~{\bf 848} 2017
	
	\bibitem{rates_abadie} J. Abadie et al., \cqg ~{\bf 27}, 173001 (2010).
	
	\bibitem{Abbott:2016nhf} 
	B.~P.~Abbott {\it et al.} [LIGO Scientific and Virgo Collaborations],
	Astrophys.\ J.\  {\bf 833}, no. 1, L1 (2016)
	doi:10.3847/2041-8205/833/1/L1
	[arXiv:1602.03842 [astro-ph.HE]].
	
	\bibitem{fermi_grb} V. Connaughton et al., \apjl ~{\bf 826} L6 (2016).
	
	\bibitem{loeb} A. Loeb, \apjl ~{819} L21 (2016).
	
	\bibitem{integral} V. Savchenko et al., \apjl ~{\bf 820}, L36 (2016)
	
	\bibitem{em_follow} L. P. Singer et al., \apjl, ~{\bf 829} L15 (2016);
	B. P. Abbott et al., \apjl ~{\bf 833} L1 (2016);
	M. Coughlin, \& C. Stubbs, Experimental Astronomy ~{\bf 42} 165 (2016);
	J. Rana, A. Singhal, B. Gadre, V. Bhalerao, \& S. Bose, \apj ~{\bf 838} 108 (2017);
	V. Srivastava, V. Bhalerao, A. P. Ravi, A. Ghosh \& S. Bose, \apj ~{\bf 838} 46 (2017);
	M. L. Chan, Y. M. Hu, C. Messenger, M. Hendry, \& I. S. Heng, \apj ~{\bf 834} 84 (2017).
	
	
	\bibitem{gw_em_cosmo}
	D. E. Holz \& S. A. Hughes, \apj ~{\bf 629}, 15 (2005);
	N. Dalal, D. E. Holz, S. A. Hughes, \& B. Jain, \prd ~{\bf 74}, 063006 (2006);
	K. G. Arun, B. R. Iyer, B. S. Sathyaprakash, S. Sinha \& C. Van Den Broeck, \prd ~{\bf 76}, 104016 (2007);
	C. Cutler \& D. E. Holz, \prd ~{\bf 80}, 104009 (2009);
	B. S. Sathyaprakash, B. F. Schutz \& C. Van Den Broeck, \cqg ~{\bf 27}
	215006 (2010);
	C. Van Den Broeck, M. Trias, B. S. Sathyaprakash, \& A. M. Sintes, \prd ~{\bf 81}, 124031 (2010);
	S. Nissanke, D. E. Holz, S. A. Hughes, N. Dalal, \& J. L. Sievers, \apj ~{\bf 725}, 496 (2010);
	S. Nissanke, D. E. Holz, N. Dalal, S. A. Hughes, J. L. Sievers \& Christopher M. Hirata, \texttt{arXiv:1307.2638}.
	
	\bibitem{markovic} D. Markovic, \prd ~{\bf 48}, 4738 (1993).
	
	\bibitem{gw_ns} D. F. Chernoff \& L. S. Finn, \apjl ~{\bf 411}, L5 (1993);
	L. S. Finn, \prd ~{\bf 53}, 2878 (1996);
	S. R. Taylor, J. R. Gair \& I. Mandel, \prd ~{\bf 85}, 023535 (2012).
	
	\bibitem{ns-mass} B. Kiziltan, A. Kottas, M. De Yoreo, \& S. E. Thorsett, \apj ~{\bf 778}, 66  (2013);
	F. Ozel, D. Psaltis, R. Narayan \& A. S. Villarreal, \apj ~{\bf 757}, 55 (2012).
	
	\bibitem{gw_tidal} C. Messenger \& J. Read, \prl ~{\bf 108}, 091101 (2012).
	
	\bibitem{walter} W. Del Pozzo, \prd ~{\bf 86}, 043011 (2012).
	
	\bibitem{hogan} C. L. MacLeod \& C. J. Hogan \prd ~{\bf 77}, 043512 (2008).
	
	\bibitem{tsv} T. D. Saini, S. K. Sethi \& V. Sahni, \prd ~{\bf 81}, 103009 (2010).
	
	\bibitem{Chen:2016tys} 
	H.~Y.~Chen and D.~E.~Holz,  arXiv:1712.06531 [astro-ph.CO].
	
	
	\bibitem{sdss} S. Alam et al., \apjs ~{\bf 219}, 12 (2015).
	
	\bibitem{Chen:2017wpg} 
	H.~Y.~Chen, D.~E.~Holz, J.~Miller, M.~Evans, S.~Vitale and J.~Creighton,
	arXiv:1709.08079 [astro-ph.CO].
	
	\bibitem{mandel_weight} I. Mandel, W. Farr \& J. Gair, LIGO Document P1600187, https://dcc.ligo.org/LIGO-P1600187/public.
	
	\bibitem{Pai:2000zt} 
	A.~Pai, S.~Dhurandhar and S.~Bose, \prd ~ {\bf 64}, 042004 (2001).
	
	\bibitem{Ghosh:2013yda} 
	S.~Ghosh and S.~Bose,
	arXiv:1308.6081 [astro-ph.HE].
	
	
	\bibitem{Bose:2011} 
	S. Bose, T. Dayanga, S. Ghosh \& D. Talukder, \cqg ~{\bf 28} 134009 (2011).
	
	
	\bibitem{broeckET} W. Zhao, C. Van Den Broeck, D. Baskaran, \& T. G. F. Li \prd ~{\bf 83}, 023005 (2011).
	
	\bibitem{oguri} M. Oguri, \prd ~{\bf 93}, 083511 (2016).
	
\end{thebibliography}
\end{document}